\documentclass[runningheads]{template/llncs}
\PassOptionsToPackage{table,svgnames,dvipsnames}{xcolor}
\usepackage[T1]{fontenc}
\usepackage{graphicx}
\usepackage{layouts}
\usepackage{booktabs} % For formal tables
\usepackage{braket}
\usepackage{bbold}
\usepackage{qcircuit}

\newcommand\identity{1\kern-0.25em\text{l}}
\usepackage{amsmath}
\usepackage{tabularray}
\usepackage{algorithm}
\usepackage{algorithmicx}
\usepackage{algpseudocode}
\usepackage{subcaption}
\usepackage{geometry}
\usepackage{supertabular}
\usepackage[hidelinks]{hyperref}
\usepackage{template/zx-calculus}
\usepackage{subcaption}
\usepackage{tikz}
\usepackage{adjustbox}
\usetikzlibrary{calc}
\usepackage{tikz/tikzit}
\usetikzlibrary{intersections}
\tikzstyle{gate}=[shape=rectangle, text height=1.5ex, text depth=0.25ex, yshift=0.5mm, fill=white, draw=black, minimum height=5mm, yshift=-0.5mm, minimum width=5mm, font={\normalsize}, tikzit category=circuit]
\tikzstyle{big gate}=[shape=rectangle, text height=1.5ex, text depth=0.25ex, yshift=0.5mm, fill=white, draw=black, minimum height=10mm, yshift=-0.5mm, minimum width=5mm, font={\normalsize}, tikzit category=circuit]
\tikzstyle{Z dot}=[inner sep=0mm, minimum size=2mm, shape=circle, draw=black, fill={rgb,255: red,221; green,255; blue,221}, tikzit category=zx]
\tikzstyle{Z phase dot}=[minimum size=5mm, font={\footnotesize\boldmath}, shape=rectangle, rounded corners=2mm, inner sep=0.2mm, outer sep=-2mm, scale=0.8, tikzit shape=circle, draw=black, fill={rgb,255: red,221; green,255; blue,221}, tikzit draw=blue, tikzit category=zx]
\tikzstyle{X dot}=[Z dot, shape=circle, draw=black, fill={rgb,255: red,255; green,136; blue,136}, tikzit category=zx]
\tikzstyle{X phase dot}=[Z phase dot, tikzit shape=circle, tikzit draw=blue, fill={rgb,255: red,255; green,136; blue,136}, font={\footnotesize\boldmath}, tikzit category=zx]
\tikzstyle{hadamard}=[fill=yellow, draw=black, shape=rectangle, inner sep=0.6mm, minimum height=1.5mm, minimum width=1.5mm, tikzit category=zx]
\tikzstyle{paulibox}=[fill={rgb,255: red,221; green,221; blue,255}, draw=black, shape=rectangle, inner sep=0.6mm, minimum height=5mm, minimum width=5mm, font={\footnotesize}, text height=1.5ex, text depth=0.25ex, tikzit category=zx]
\tikzstyle{vertex}=[inner sep=0mm, minimum size=1mm, shape=circle, draw=black, fill=black, tikzit category=misc]
\tikzstyle{vertex set}=[inner sep=0mm, minimum size=1mm, shape=circle, draw=black, fill=white, font={\footnotesize\boldmath}, tikzit category=misc]
\tikzstyle{small black dot}=[fill=black, draw=black, shape=circle, inner sep=0pt, minimum width=1.2mm, tikzit category=circuit]
\tikzstyle{cnot ctrl}=[fill=black, draw=black, shape=circle, inner sep=0pt, minimum width=1.2mm, tikzit category=circuit]
\tikzstyle{cnot targ}=[fill=white, draw=white, shape=circle, tikzit category=circuit, label={center:$\oplus$}, inner sep=0pt, minimum width=2.1mm, tikzit fill={rgb,255: red,102; green,204; blue,255}, tikzit draw=black]
\tikzstyle{ket}=[fill=white, draw=black, shape=regular polygon, regular polygon sides=3, regular polygon rotate=-30, scale=0.7, inner sep=1pt, tikzit category=circuit, tikzit shape=rectangle, tikzit fill=green]
\tikzstyle{bra}=[fill=white, draw=black, shape=regular polygon, regular polygon sides=3, regular polygon rotate=30, scale=0.7, inner sep=1pt, tikzit category=circuit, tikzit shape=rectangle, tikzit fill=red]
\tikzstyle{scalar}=[shape=rectangle, text height=1.5ex, text depth=0.25ex, yshift=0.5mm, fill=white, draw=black, minimum height=5mm, yshift=-0.5mm, minimum width=5mm, font={\normalsize}]
\tikzstyle{clabel}=[fill=white, draw=none, shape=rectangle, tikzit fill={rgb,255: red,56; green,255; blue,242}, font={\footnotesize}, inner sep=1pt, tikzit category=labels]
\tikzstyle{empty diagram}=[draw={gray!40!white}, dashed, shape=rectangle, minimum width=1cm, minimum height=1cm, tikzit category=misc]

\tikzstyle{hadamard edge}=[-, dashed, dash pattern=on 2pt off 0.5pt, thick, draw={rgb,255: red,68; green,136; blue,255}]
\tikzstyle{box edge}=[-, dashed, dash pattern=on 2pt off 0.5pt, thick, draw={rgb,255: red,203; green,192; blue,225}]
\tikzstyle{brace edge}=[-, tikzit draw=blue, decorate, decoration={brace,amplitude=1mm,raise=-1mm}]
\tikzstyle{diredge}=[->]
\tikzstyle{double edge}=[-, double, shorten <=-1mm, shorten >=-1mm, double distance=2pt]
\tikzstyle{gray edge}=[-, {gray!70!white}, thick]
\tikzstyle{pointer edge}=[->, very thick, gray]
\tikzstyle{boldedge}=[-, line width=1.6pt, shorten <=-0.17mm, shorten >=-0.17mm]

\definecolor{luxembourg blue}{HTML}{00a4dd}                                                                                                                          
\definecolor{luxembourg red}{HTML}{e31b1d}

\usetikzlibrary{shapes,arrows,intersections,patterns}
\usetikzlibrary{calc, arrows.meta, intersections, patterns, positioning, shapes.misc, fadings, through,decorations.pathreplacing}
\usetikzlibrary{matrix,fit,calc,trees,positioning,arrows,chains,shapes.geometric,shapes}
\newcolumntype{C}[1]{>{\centering\let\newline\\\arraybackslash\hspace{0pt}}m{#1}}

\DeclareGraphicsExtensions{.pdf,.jpeg,.png}

\usepackage{longtable}
\usepackage{graphicx}
\newsavebox\CBox
\def\textBF#1{\sbox\CBox{#1}\resizebox{\wd\CBox}{\ht\CBox}{\textbf{#1}}}
\usepackage{url}
\usepackage{breakurl}

\begin{document}
\title{Exhaustive Search for Quantum Circuit
Optimization using ZX Calculus}
\titlerunning{Exhaustive Search for Quantum Circuit
Optimization using ZX Calculus}
\author{Tobias~Fischbach\inst{1}\orcidID{0009-001-2535-2577} \and
  Pierre~Talbot\inst{1}\orcidID{0000-0001-9202-4541} \and~Pascal~Bouvry\inst{1}\orcidID{0000-0001-9338-2834}}

\authorrunning{T. Fischbach et al.}

\institute{Department of Computer Science, University of Luxembourg,
  Esch-sur-Alzette, Luxembourg
\email{firstname.lastname@uni.lu}}
\maketitle              % typeset the header of the contribution
\begin{abstract}
  Quantum computers allow a near-exponential speed-up for specific
  applications when compared to classical computers.
  Despite recent advances in the hardware of quantum computers, their
  practical usage is still severely limited due to a restricted
  number of available physical qubits and quantum gates, short coherence time,
  and high error rates.
  This paper lays the foundation towards a metric independent
  approach to quantum
  circuit optimization based on exhaustive search algorithms.
  This work uses depth-first search and iterative deepening depth-first search.
  We rely on ZX calculus to represent and optimize quantum circuits
  through the minimization of a given metric (e.g. the T-gate and edge count).
  ZX calculus formally guarantees that the semantics of the original
  circuit is preserved.
  As ZX calculus is a non-terminating rewriting system, we utilise a
  novel set of pruning rules to ensure termination while still
  obtaining high-quality solutions.
  We provide the first formalization of quantum circuit optimization
  using ZX calculus and exhaustive search.
  We extensively benchmark our approach on
  100 standard quantum circuits.
  Finally, our implementation is integrated in the well-known libraries
  PyZX and Qiskit as a compiler pass to ensure applicability of our results.
   % final period required by SCOPUS
\end{abstract}

\setlength{\textfloatsep}{0pt}
\setlength{\floatsep}{0pt}
\setlength{\intextsep}{05pt}

\section{Introduction}
\label{sec:introduction}
Quantum computers allow a near-exponential speed-up for specific
applications when compared to classical computers. Typical
examples that are benefitted by quantum computing include the simulation of quantum systems, solving
combinatorial problems, performing machine learning and breaking
cryptography~\cite{hassijaForthcomingApplicationsQuantum2020}.
However, current quantum computers lack the resources to address
complex real-world problems. These restrictions concern the number of
available physical qubits and quantum gates, high error rates, and a short
coherence time. Quantum error correction aims to mitigate these challenges at the cost of a higher resource demand~\cite{campbellRoadsFaulttolerantUniversal2017}.

Similarly to classical computing, quantum circuits describe quantum
programs within the quantum gate model. These circuits are
independent of an underlying architecture and allow universal
computation~\cite{chi-chihyaoQuantumCircuitComplexity1993}. Users
typically use Clifford gates with the T-gate as the chosen universal
gate set because it can be efficiently simulated on classical
computers~\cite{aaronsonImprovedSimulationStabilizer2004}.
However, different types of gates require varying amounts of resources,
with the T-gate requiring more physical qubits and error correction code than
Clifford gates to be implemented in a quantum device~\cite{fowlerSurfaceCodesPractical2012}.

Inherent limitations in quantum architecture are the number of
available physical qubits and quantum gates, as well as a short coherence time.
These limitations can be addressed by architecture-dependent
optimization that improves the mapping of a quantum circuit onto a
specific quantum hardware.
Architecture-dependent quantum circuit optimization can be treated as a classical
optimization problem that can be solved
exactly~\cite{venturelliCompilingQuantumCircuits2018}.
Other methods include heuristics and deep
learning~\cite{khairyReinforcementLearningBasedVariationalQuantum2019,foselQuantumCircuitOptimization2021}.

In this paper, we target \textit{architecture-independent
optimization} which aims to simplify a quantum circuit by reducing
general and common limitation factors across architectures such as
the number of quantum gates, logical qubits and the circuit
depth.
Despite the existence of infinite universal gate sets, common methods employ gate commutation
rules~\cite{itokoOptimizationQuantumCircuit2020} and circuit
simplification~\cite{maslovQuantumCircuitSimplification2008} for frequently used universal gate sets.
These heuristic approaches suffer from several drawbacks.
First, it is necessary to prove that each new simplification rule is
correct, to make sure that the semantics of the original circuit is preserved. 
Second, there is no guarantee of optimality, even for small circuits.
And finally, the heuristics are tailored to optimize one particular
objective and must be redesigned when the requirements change.
To tackle these challenges, several optimization methods based on ZX
calculus (Section~\ref{sec:preliminaries}) recently
emerged~\cite{duncanGraphtheoreticSimplificationQuantum2020,nagele}.

ZX calculus is a universal, compact and complete rewriting
system~\cite{coeckeInteractingQuantumObservables2008a,coeckePicturingQuantumProcesses2017}.
An object in ZX calculus is depicted graphically and called a
\textit{ZX diagram}.
Quantum circuits and ZX diagrams both represent a linear map between qubits.
ZX calculus is \textit{universal} because every quantum circuit can
be converted to a ZX diagram.
It is \textit{complete} because applying any rule preserves
semantics, which means that the linear map of qubits remains
unchanged~\cite{backensZXcalculusCompleteStabilizer2014,backensMakingStabilizerZXcalculus2015,jeandelCompleteAxiomatisationZXCalculus2018}.
Finally, ZX calculus is \textit{compact} because it consists only of
two generators and eight rules.
A ZX diagram can be converted back to a quantum circuit, which is a
non-trivial process known as the \textit{circuit extraction
problem}~\cite{debeaudrapCircuitExtractionZXdiagrams2022}.
% The power of ZX calculus lies in its abstraction of the quantum
% circuit notation.
% Instead of many gate commutation rules in the quantum circuit
% notation, two generators and a compact set of 9 rules show in
% Figure~\ref{fig:rules} and~\ref{fig:local-complementation}.

However, some characteristics of ZX calculus make the design of an
optimization algorithm challenging:
\begin{itemize}
  \item \textbf{High memory requirements}: Real-world quantum
    circuits result in large ZX diagrams with high memory
    requirements for every state.
  \item \textbf{Non-terminating}: Infinite rewriting sequences exist.
  \item \textbf{Failed states}: The extraction of ZX diagrams to
    quantum circuit might fail (or be prohibitively long) with
    current algorithms. Checking if a diagram is extractable is
    time consuming and prevents us from exploring a large number of nodes.
\end{itemize}

In light of these challenges, we contribute a proof of concept
to quantum circuit optimization by employing \textit{iterative
deepening depth-first search} (IDDFS) to systematically explore the
rewritten ZX diagrams.
IDDFS is a well-known state-space search strategy, which is simple,
memory efficient and provides a good trade-off between exploration
and exploitation~\cite{korfDepthfirstIterativedeepeningOptimal1985}.
Our approach is general in the sense that the same search strategy
can be employed
to optimize different metrics. A metric can be defined on the basis of
characteristics of the ZX diagram
or its corresponding quantum circuit.
In particular, we aim to find a ZX diagram that minimizes the T-gate
count due to its high impact on the practicability of current quantum architectures.
Furthermore, we demonstrate the metric independence of our approach
by optimizing the edge count of the ZX diagram.
In sum, the contributions of this paper are as follows:
\begin{enumerate}
  \item A formal description of ZX diagram optimization and the
    first state-space search algorithm applied to ZX diagram
    optimization (Section~\ref{sec:dfs-for-zx}).
  \item Proof of concept implementation that is extensively
    benchmarked on 100 standard
    quantum circuits that is equating the state-of-the-art full reduce algorithm on $89\%$ of the
    circuits within 1.5 hours.
    (Section~\ref{sec:t-gate-reduction}).
  \item A PyZX and Qiskit based transpiler pass to support the
    practical adoption of our
    results.
\end{enumerate}

\section{Preliminaries}
\label{sec:preliminaries}

\subsection{ZX Calculus}
\label{sec:zx_calculus}

ZX calculus, introduced by Coecke and Duncan in 2008, is a universal
framework for diagrammatic reasoning between linear maps of
qubits~\cite{coeckeInteractingQuantumObservables2008,coeckePicturingQuantumProcesses2017}.
It provides a complete set of sound semantic-preserving rewriting
rules, even for arbitrary real
phases~\cite{backensZXcalculusCompleteStabilizer2014,backensMakingStabilizerZXcalculus2015,jeandelCompleteAxiomatisationZXCalculus2018,jeandelDiagrammaticReasoningClifford2018}.
Completeness signifies that the linear map of a ZX diagram remains
unchanged after modifications through a rewriting rule. ZX calculus
is universal and allows users to represent any quantum circuit as a
ZX diagram. The elementary building blocks of quantum circuits are
quantum gates and wires. Analogously, the elementary building blocks
in ZX calculus---called \textit{generators}---are spiders, wires,
swap and Bell states.

A spider is a tensor which operates on qubits in either the Z-basis
$\left\{\ket{0}, \ket{1} \right\}$ (green) or X-basis
$\left\{\ket{-}, \ket{+} \right\}$ (red).  Spiders possess $n$
inputs, $m$ outputs and carry a phase $\alpha$.
The linear map of a spider for phase $\alpha = 0$ results in the
identity matrix $\underset{2^{m}\times 2^{n}}{\identity}$ and
therefore acts as a wire.
Spiders with phases that are multiples of $\frac{\pi}{2}$ can
implement all Clifford gates. The T-gate corresponds to a Z-spider
with a phase of $\frac{\pi}{4}$. Clifford gates and the T-gate form a
universal gate set together.

%\noindent{
% \begin{center}
\begin{figure}[t]
  %\setlength\arraycolsep{1.5pt}
  %\resizebox{\textwidth}{!}{
  \begin{tabular}{C{3cm}C{3.5cm}C{3.5cm}C{3cm}}
    \textbf{ID-removal} & \textbf{Fusion} & \textbf{Colour change}
    & \textbf{Bialgebra}\\
    \begin{tikzpicture}
	\begin{pgfonlayer}{nodelayer}
		\node [style=none] (31) at (-12, 4.375) {$\overset{(i1)}{=}$};
		\node [style=none] (32) at (-12, 3.6) {$\overset{(i2)}{=}$};

		\node [style=none] (36) at (-11, 3) {};
		\node [style=hadamard] (35) at (-11.75, 3) {};
		\node [style=hadamard] (41) at (-12.25, 3) {};
		\node [style=none] (37) at (-13, 3) {};

		\node [style=none] (38) at (-11, 5) {};
		\node [style=Z dot] (39) at (-12, 5) {};
		\node [style=none] (40) at (-13, 5) {};

		\node [style=none] (33) at (-11, 4) {};
		\node [style=none] (34) at (-13, 4) {};
	\end{pgfonlayer}
	\begin{pgfonlayer}{edgelayer}
		\draw (36.center) to (35);
		\draw (38.center) to (39);
		\draw (39) to (40.center);
		\draw (33.center) to (34.center);
		\draw (35) to (41);
		\draw (41) to (37.center);
	\end{pgfonlayer}
\end{tikzpicture} & \begin{tikzpicture}
	\begin{pgfonlayer}{nodelayer}
		\node [style=none] (11) at (-10.75, 3.5) {$\overset{(f)}{=}$};
		\node [style=Z phase dot] (12) at (-12, 3.5) {$\beta$};
		\node [style=none] (13) at (-11, 2.5) {};
		\node [style=none] (14) at (-12.75, 2.5) {};
		\node [style=Z phase dot] (15) at (-13.25, 4.25) {$\alpha$};
		\node [style=none] (16) at (-12.5, 5) {};
		\node [style=none] (17) at (-14, 5) {};
		\node [style=none] (18) at (-14, 2.5) {};
		\node [style=Z phase dot] (19) at (-9, 3.75) {$\alpha+\beta$};
		\node [style=none] (20) at (-10, 5) {};
		\node [style=none] (21) at (-10, 2.5) {};
		\node [style=none] (22) at (-8, 2.5) {};
		\node [style=none] (23) at (-8, 5) {};
		\node [style=none] (24) at (-13, 3.5) {\small$\ldots$};
		\node [style=none] (25) at (-13.25, 5) {\small$\ldots$};
		\node [style=none] (26) at (-9, 4.75) {\small$\ldots$};
		\node [style=none] (27) at (-9, 3) {\small$\ldots$};
		\node [style=none] (28) at (-12, 2.5) {\small$\ldots$};
		\node [style=none] (29) at (-11, 5) {};
		\node [style=none] (30) at (-12, 4.25) {\small$\ldots$};
	\end{pgfonlayer}
	\begin{pgfonlayer}{edgelayer}

		\draw [in=90, out=-5, looseness=0.75] (12) to (13.center);
		\draw [in=90, out=-175, looseness=0.75] (12) to (14.center);

		\draw [in=185, out=90, looseness=0.75] (18.center) to (15);

		\draw [in=-90, out=175, looseness=0.75] (15) to (17.center);
		\draw [in=-90, out=5, looseness=0.75] (15) to (16.center);

		\draw [in=150, out=-15] (15) to (12);

		\draw [in=90, out=-5, looseness=0.75] (19) to (22.center);
		\draw [in=90, out=-175, looseness=0.75] (19) to (21.center);
		\draw [in=-90, out=5, looseness=0.75] (19) to (23.center);
		\draw [in=-90, out=175, looseness=0.75] (19) to (20.center);

		\draw [in=-90, out=5] (12) to (29.center);
	\end{pgfonlayer}
\end{tikzpicture} &\begin{tikzpicture}
	\begin{pgfonlayer}{nodelayer}
		\node [style=X phase dot] (71) at (-13.75, 3.5) {$\alpha$};
		\node [style=none] (76) at (-12.75, 3.75) {$\vdots$};
		\node [style=none] (77) at (-14.75, 3.75) {$\vdots$};
		\node [style=none] (78) at (-12.25, 3.5) {$\overset{(h)}{=}$};

		\node [style=none] (72) at (-12.75, 4.5) {};
		\node [style=none] (73) at (-14.75, 4.5) {};
		\node [style=none] (74) at (-14.75, 2.5) {};
		\node [style=none] (75) at (-12.75, 2.5) {};

		\node [style=none] (80) at (-11.75, 4.5) {};
		\node [style=none] (83) at (-11.75, 2.5) {};
		\node [style=none] (81) at (-9.75, 4.5) {};
		\node [style=none] (82) at (-9.75, 2.5) {};

		\node [style=Z phase dot] (79) at (-10.75, 3.5) {$\alpha$};
		\node [style=none] (84) at (-11.75, 3.75) {$\vdots$};
		\node [style=none] (85) at (-9.75, 3.75) {$\vdots$};

		\node [style=hadamard] (86) at (-11.75, 4.5) {};
		\node [style=hadamard] (87) at (-9.75, 2.5) {};
		\node [style=hadamard] (88) at (-9.75, 4.5) {};
		\node [style=hadamard] (89) at (-11.75, 2.5) {};

		\node [style=none] (90) at (-12.125, 4.5) {};
		\node [style=none] (91) at (-9.375, 2.5) {};
		\node [style=none] (92) at (-9.375, 4.5) {};
		\node [style=none] (93) at (-12.125, 2.5) {};

		\node [style=none] (10) at (-15.125, 4.5) {};
		\node [style=none] (11) at (-12.375, 2.5) {};
		\node [style=none] (12) at (-12.375, 4.5) {};
		\node [style=none] (13) at (-15.125, 2.5) {};
	\end{pgfonlayer}
	\begin{pgfonlayer}{edgelayer}
		\draw [in=180, out=85, looseness=0.75] (71) to (72.center);
		\draw [in=0, out=95] (71) to (73.center);
		\draw [in=0, out=-95, looseness=0.75] (71) to (74.center);
		\draw [in=180, out=-85, looseness=0.75] (71) to (75.center);

		\draw [in=0, out=95, looseness=0.75] (79) to (80.center);
		\draw [in=0, out=-95, looseness=0.75] (79) to (83.center);
		\draw [in=180, out=85, looseness=0.75] (79) to (81.center);
		\draw [in=180, out=-85, looseness=0.75] (79) to (82.center);

		\draw (86) to (90.center);
		\draw (87) to (91.center);
		\draw (88) to (92.center);
		\draw (89) to (93.center);

		\draw (75.center) to (11.center);
		\draw (72.center) to (12.center);
		\draw (73.center) to (10.center);
		\draw (74.center) to (13.center);
	\end{pgfonlayer}
\end{tikzpicture}
    & \begin{tikzpicture}
	\begin{pgfonlayer}{nodelayer}
		\node [style=none] (42) at (-11, 3.5) {$\overset{(b)}{=}$};
		\node [style=none] (43) at (-11.25, 4.5) {};
		\node [style=none] (44) at (-11.25, 2.5) {};
		\node [style=Z dot] (45) at (-12.25, 3.5) {};
		\node [style=X dot] (46) at (-13, 3.5) {};
		\node [style=none] (47) at (-14, 4.5) {};
		\node [style=none] (48) at (-14, 2.5) {};
		\node [style=none] (49) at (-10.25, 4.5) {};
		\node [style=none] (50) at (-10.25, 2.5) {};
		\node [style=Z dot] (51) at (-9.75, 4.5) {};
		\node [style=Z dot] (52) at (-9.75, 2.5) {};
		\node [style=X dot] (53) at (-8.5, 2.5) {};
		\node [style=X dot] (54) at (-8.5, 4.5) {};
		\node [style=none] (55) at (-8, 4.5) {};
		\node [style=none] (56) at (-8, 2.5) {};
	\end{pgfonlayer}
	\begin{pgfonlayer}{edgelayer}
		\draw [in=95, out=0] (47.center) to (46);
		\draw [in=0, out=-95] (46) to (48.center);
		\draw (45) to (46);
		\draw [in=180, out=85] (45) to (43.center);
		\draw [in=180, out=-85] (45) to (44.center);
		\draw (50.center) to (52);
		\draw (52) to (54);
		\draw (49.center) to (51);
		\draw (51) to (53);
		\draw (53) to (56.center);
		\draw (54) to (55.center);
		\draw (51) to (54);
		\draw (52) to (53);
	\end{pgfonlayer}
\end{tikzpicture}\\[-3em]
    \textbf{Copy} &  \textbf{Pi-Copy} & \textbf{Euler} &
    \textbf{Local comp.}          \\
    \begin{tikzpicture}
	\begin{pgfonlayer}{nodelayer}
		\node [style=none] (0) at (-11, 2) {$\overset{(c)}{=}$};
		\node [style=none] (1) at (-12, 2.75) {};
		\node [style=none] (2) at (-12, 2) {\vdots};
		\node [style=none] (3) at (-12, 0.75) {};
		\node [style=Z phase dot] (4) at (-13, 1.75) {$\alpha$};
		\node [style=X dot] (5) at (-14, 1.75) {};
		\node [style=none] (6) at (-9.25, 2.75) {};
		\node [style=none] (7) at (-10, 2) {\vdots};
		\node [style=none] (8) at (-9.25, 0.75) {};
		\node [style=X dot] (9) at (-10, 2.75) {};
		\node [style=X dot] (10) at (-10, 0.75) {};
	\end{pgfonlayer}
	\begin{pgfonlayer}{edgelayer}
		\draw (5) to (4);
		\draw [in=-180, out=85] (4) to (1.center);
		\draw [in=-180, out=-85] (4) to (3.center);
		\draw (9) to (6.center);
		\draw (10) to (8.center);
	\end{pgfonlayer}
\end{tikzpicture} & \begin{tikzpicture}
	\begin{pgfonlayer}{nodelayer}
		\node [style=none] (57) at (-10.75, 2) {$\overset{(\pi)}{=}$};
		\node [style=none] (58) at (-11.5, 3) {};
		\node [style=none] (59) at (-11.75, 2.25) {\vdots};
		\node [style=none] (60) at (-11.5, 1) {};
		\node [style=Z phase dot] (61) at (-12.75, 2) {$\alpha$};
		\node [style=X phase dot] (62) at (-13.75, 2) {$\pi$};
		\node [style=none] (63) at (-14.5, 2) {};
		\node [style=none] (64) at (-10, 2) {};
		\node [style=Z phase dot] (65) at (-9.25, 2) {$-\alpha$};
		\node [style=none] (66) at (-7.625, 3) {};
		\node [style=none] (67) at (-8.25, 2.25) {\vdots};
		\node [style=none] (68) at (-7.625, 1) {};
		\node [style=X phase dot] (69) at (-8.25, 3) {$\pi$};
		\node [style=X phase dot] (70) at (-8.25, 1) {$\pi$};
	\end{pgfonlayer}
	\begin{pgfonlayer}{edgelayer}
		\draw (63.center) to (62);
		\draw (62) to (61);
		\draw [in=-180, out=85] (61) to (58.center);
		\draw [in=-180, out=-85] (61) to (60.center);
		\draw (64.center) to (65);
		\draw [in=-180, out=85] (65) to (69);
		\draw (69) to (66.center);
		\draw [in=180, out=-85] (65) to (70);
		\draw (70) to (68.center);
	\end{pgfonlayer}
\end{tikzpicture} & \begin{tikzpicture}
	\begin{pgfonlayer}{nodelayer}
		\node [style=hadamard] (90) at (-11.75, 2.75) {};
		\node [style=none] (91) at (-10.75, 2.75) {};
		\node [style=none] (92) at (-12.75, 2.75) {};
		\node [style=none] (93) at (-10.0, 0.75) {};
		\node [style=none] (95) at (-13.5, 0.75) {};
    \node [style=X phase dot] (96) at (-11.75, 0.75) {$\frac{\pi}{2}$};
		\node [style=Z phase dot] (97) at (-10.75, 0.75) {$\frac{\pi}{2}$};
		\node [style=Z phase dot] (98) at (-12.75, 0.75) {$\frac{\pi}{2}$};
		\node [style=none] (99) at (-11.75, 1.75) {$\overset{(hd)}{=}$};
	\end{pgfonlayer}
	\begin{pgfonlayer}{edgelayer}
		\draw (91.center) to (90);
		\draw (90) to (92.center);
		\draw (93.center) to (97);
		\draw (97) to (96);
		\draw (96) to (98);
		\draw (95.center) to (98);
	\end{pgfonlayer}
\end{tikzpicture} &
    \begin{tikzpicture}
	\begin{pgfonlayer}{nodelayer}
    \node [style=Z dot] (0) at ((-1.5, 2.75) {};
		\node [style=Z dot, draw=luxembourg blue, ultra thick] (1) at (-3.5, 2.75) {};
		\node [style=Z dot] (2) at (-3.5, 1) {};
		\node [style=Z dot] (3) at (-1.5, 1) {};
    \node [style=none] (18) at (0, 2) {$\overset{(lc)}{=}$};
		\node [style=Z dot] (19) at (3.5, 2.75) {};
		\node [style=Z dot] (21) at (1.25, 1) {};
		\node [style=Z dot] (29) at (1.25, 2.75) {};
		\node [style=Z dot] (22) at (3.5, 1) {};
		\node [style=none] (23) at (-3.5, 3.5) {$a$};
		\node [style=none] (30) at (-1.5, 3.5) {$b$};
		\node [style=none] (23) at (-3.5, 0.5) {$c$};
		\node [style=none] (30) at (-1.5, 0.5) {$d$};
		\node [style=none] (23) at (1.25, 3.5) {$a$};
		\node [style=none] (30) at (3.5, 3.5) {$b$};
		\node [style=none] (23) at (1.25, 0.5) {$c$};
		\node [style=none] (30) at (3.5, 0.5) {$d$};
		\node [style=none] (25) at (-4, 3.75) {};
		\node [style=none] (26) at (-4, 2.25) {};
		\node [style=none] (27) at (-3, 3.75) {};
		\node [style=none] (28) at (-3, 2.25) {};

	\end{pgfonlayer}
	\begin{pgfonlayer}{edgelayer}
		\draw  (1) to (0);
		\draw  (1) to (2);
		\draw  (0) to (3);
		\draw  (1) to (3);
    \draw (19) to (21);
    \draw (21) to (29);
    \draw (21) to (29);
    \draw (29) to (22);
    \draw (22) to (21);
    %\draw (19) to (22);
    \draw (19) to (29);
		%\draw [style=hadamard edge] (21) to (19);
		%\draw [style=hadamard edge] (2) to (3);
		%\draw [style=hadamard edge] (19) to (22);
		%\draw [style=hadamard edge] (1) to (0);
		%\draw [style=hadamard edge] (1) to (2);
		%\draw [style=hadamard edge] (1) to (3);
		%\draw [style=hadamard edge] (21) to (19);
		%\draw [style=hadamard edge] (2) to (3);
		%\draw [style=hadamard edge] (19) to (22);
		%\draw [style=gray edge] (25.center) to (27.center);
		%\draw [style=gray edge] (27.center) to (28.center);
		%\draw [style=gray edge] (28.center) to (26.center);
		%\draw [style=gray edge] (26.center) to (25.center);
	\end{pgfonlayer}
\end{tikzpicture} \\
  \end{tabular}
  \caption{The basic rewriting rules of ZX calculus.}
  \label{fig:rules}
\end{figure}
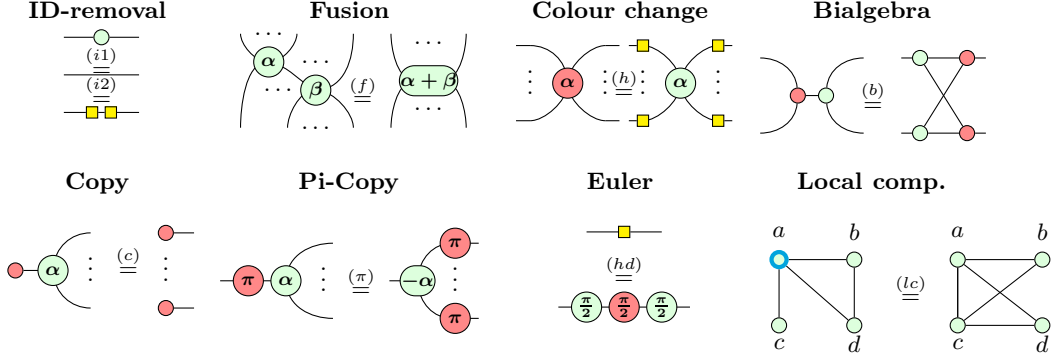
%  \end{center}
%}

\begin{center}
  \begin{tabular}{lcl}
    \begin{ZX}  \leftManyDots{n} \zxZ{\alpha} \rightManyDots{m}
    \end{ZX}  & $=$ & $\ket{0, ...,0}\bra{0,...,0}
    +e^{i\alpha}\ket{1, ...,1}\bra{1,...,1}$ \\
    \begin{ZX}  \leftManyDots{n} \zxX{\alpha} \rightManyDots{m}
    \end{ZX} & $=$ & $\ket{+, ...,+}\bra{+,...,+} +e^{i\alpha}\ket{-,
    ...,-}\bra{-,...,-}$
  \end{tabular}
\end{center}

Wires connect the outputs of one spider with the inputs of other
spiders. The identity matrix $\identity$ implements the linear map of wires.
\begin{center}
  \begin{tabular}{lcl}
    \begin{ZX}[ampersand replacement=\&] \zxN{} \rar \&[\zxWCol] \zxN{}
    \end{ZX} & $=$ & $\ket{0}\bra{0}  + \ket{1}\bra{1}$ \\
  \end{tabular}
\end{center}
A yellow box connected by wires indicates a Hadamard generator. Euler
decomposition, known as the Hadamard rule (hd), splits a Hadamard generator
into a sequence of Z and
X spiders~\cite{duncanGraphStatesNecessity2009}.
\begin{center}
  \begin{tabular}{lcl}
    \begin{ZX}
      \zxN{} \rar &[\zxwCol] \zxH{} \rar &[\zxwCol] \zxN{}
    \end{ZX} & $=$ &
    \begin{ZX}
      \zxN{} \ar[r] & \zxFracZ{\pi}{2} \ar[r] & \zxFracX{\pi}{2}
      \ar[r] & \zxFracZ{\pi}{2} \ar[r] & \zxN{}
    \end{ZX} \\
  \end{tabular}
\end{center}
The swap generator swaps the spiders on a wire and implements the
identical linear map as the swap gate in the quantum circuit notation.
\begin{center}
  \begin{tabular}{lcl}
    \begin{ZX}
      \zxNoneDouble|{} \ar[r,s,start anchor=north,end anchor=south]
      \ar[r,s,start
      anchor=south,end anchor=north] &[\zxWCol] \zxNoneDouble|{}
    \end{ZX} & $=$ & $\ket{00}\bra{00}  + \ket{01}\bra{10}+
    \ket{10}\bra{01}+ \ket{11}\bra{11}$ \\
  \end{tabular}
\end{center}
In ZX calculus, bent wires depict the Bell state and the Bell effect and
are known as cup and cap.
\begin{center}
  \begin{tabular}{lcl}
    \begin{ZX}
      \zxNone{} \ar[d,C] \\[\zxWRow]
      \zxNone{}
    \end{ZX}  & $=$ & $\ket{00} + \ket{11}$ \\
    \begin{ZX}
      \zxNone{} \ar[d,C-] \\[\zxWRow]
      \zxNone{}
    \end{ZX} & $=$ & $\bra{00} + \bra{11}$  \\
  \end{tabular}
\end{center}

A typical ZX diagram consists of many connected spiders and Hadamard
generators. Matrix multiplication composes the linear map of
sequentially connected spiders. The tensor product composes the linear
map between non-sequential connected spiders and Hadamards, meaning that
generators are parallel to each other.

\textit{Only topology matters} is an important concept in ZX
calculus. It states that the linear map between qubits of a ZX
diagram remains unchanged as long as its connectivity stays the same.
As a consequence, bending wires (e.g. cups and caps) and
moving spiders do not change the ZX
diagram~\cite{coeckePicturingQuantumProcesses2017}.

\subsection{Rewriting Rules}
\label{sec:zxrules}

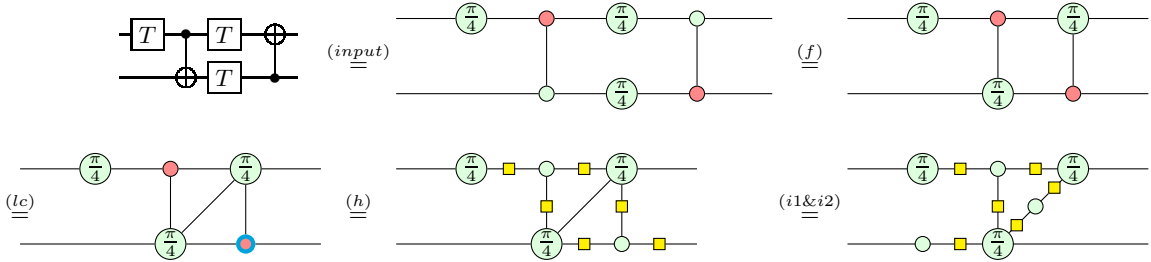
\begin{figure}[h!]
  %\begin{center}
  %\resizebox{\textwidth}{!}{
  \begin{tikzpicture}
	\begin{pgfonlayer}{nodelayer}
		\node [style=none] (10) at (-2, 2) {};
		\node [style=none] (11) at (-2, 0) {};
		\node [style=none] (12) at (8, 2) {};
		\node [style=none] (13) at (8, 0) {};
		\node [style=none] (14) at (9,1) {$\overset{(f)}{=}$};
    \node [style=none] (15) at (-3,1) {$\overset{(input)}{=}$};

    \node (100) at (-7,1) {
              
          $\Qcircuit @C=0.5em @R=.5em @!R {
            & \gate{T}  & \ctrl{1}  & \gate{T}  & \qw & \targ    & \qw \\
            & \qw       & \targ     & \gate{T}  & \qw & \ctrl{-1} & \qw}$

    };

		\node [style=none] (20) at (10, 2) {};
		\node [style=none] (21) at (10, 0) {};
		\node [style=none] (22) at (18, 2) {};
		\node [style=none] (23) at (18, 0) {};

		\node [style=none] (25) at (-12, -2) {};
		\node [style=none] (26) at (-12, -4) {};
		\node [style=none] (27) at (-4, -2) {};
		\node [style=none] (28) at (-4, -4) {};

    \node [style=Z dot] (1) at (0,2) {$\frac{\pi}{4}$};
    \node [style=X dot] (2) at (2,2) {};
    \node [style=Z dot] (3) at (4,2) {$\frac{\pi}{4}$};
    \node [style=Z dot] (4) at (6,2) {};
    \node [style=Z dot] (5) at (2,0) {};
    \node [style=Z dot] (6) at (4,0) {$\frac{\pi}{4}$};
    \node [style=X dot] (7) at (6,0) {};

    \node [style=Z dot] (31) at (12,2) {$\frac{\pi}{4}$};
    \node [style=X dot] (32) at (14,2) {};
    \node [style=Z dot] (33) at (16,2) {$\frac{\pi}{4}$};
    \node [style=Z dot] (34) at (14,0) {$\frac{\pi}{4}$};
    \node [style=X dot] (35) at (16,0) {};

    \node [style=Z dot] (41) at (-10,-2) {$\frac{\pi}{4}$};
    \node [style=X dot] (42) at (-8,-2) {};
    \node [style=Z dot] (43) at (-6,-2) {$\frac{\pi}{4}$};
    \node [style=Z dot] (44) at (-8,-4) {$\frac{\pi}{4}$};
    \node [style=X dot,draw=luxembourg blue, ultra thick] (45) at (-6,-4) {};
		\node [style=none] (24) at (-12,-3) {$\overset{(lc)}{=}$};

    \node [style=Z dot] (51) at (0,-2) {$\frac{\pi}{4}$};
    \node [style=Z dot] (52) at (2,-2) {};
    \node [style=hadamard] (53) at (1,-2) {};
    \node [style=Z dot] (54) at (4,-2) {$\frac{\pi}{4}$};
    \node [style=hadamard] (55) at (3,-2) {};
    \node [style=hadamard] (56) at (2,-3) {};
    \node [style=Z dot] (58) at (2,-4) {$\frac{\pi}{4}$};
    \node [style=hadamard] (60) at (3,-4) {};
    \node [style=Z dot] (61) at (4,-4) {};
    \node [style=hadamard] (62) at (4,-3) {};
    \node [style=hadamard] (63) at (5,-4) {};

		\node [style=none] (64) at (-2, -2) {};
		\node [style=none] (65) at (-2, -4) {};
		\node [style=none] (66) at (6, -2) {};
		\node [style=none] (67) at (6, -4) {};
		\node [style=none] (68) at (-3,-3) {$\overset{(h)}{=}$};
    \node [style=none] (68) at (9,-3) {$\overset{(i1 \& i2)}{=}$};

		\node [style=none] (70) at (10, -2) {};
		\node [style=none] (83) at (10, -4) {};
    \node [style=Z dot] (71) at (12,-2) {$\frac{\pi}{4}$};
    \node [style=hadamard] (72) at (13,-2) {};
    \node [style=Z dot] (73) at (14,-2) {};
    \node [style=hadamard] (74) at (15,-2) {};
    \node [style=Z dot] (75) at (16,-2) {$\frac{\pi}{4}$};
		\node [style=none] (76) at (18, -2) {};

    \node [style=Z dot] (77) at (12,-4) {};
    \node [style=hadamard] (78) at (13,-4) {};
    \node [style=Z dot] (79) at (14,-4) {$\frac{\pi}{4}$};
    \node [style=hadamard] (80) at (14.5,-3.5) {};
    \node [style=Z dot] (81) at (15,-3) {};
    \node [style=hadamard] (82) at (15.5,-2.5) {};
		\node [style=none] (84) at (18, -4) {};
    \node [style=hadamard] (85) at (14,-3) {};

	\end{pgfonlayer}
	\begin{pgfonlayer}{edgelayer}
    \draw (10.center) to (1);
    \draw (1) to (2);
    \draw (2) to (3);
    \draw (3) to (4);
    \draw (4) to (12.center);
    \draw (11.center) to (5);
    \draw (5) to (6);
    \draw (6) to (7);
    \draw (7) to (13.center);
    \draw (2) to (5);
    \draw (4) to (7);

    \draw (20.center) to (31);
    \draw (31) to (32);
    \draw (32) to (33);
    \draw (33) to (22.center);
    \draw (21.center) to (34);
    \draw (34) to (35);
    \draw (35) to (23.center);
    \draw (32) to (34);
    \draw (33) to (35);

    \draw (25.center) to (41);
    \draw (41) to (42);
    \draw (42) to (43);
    \draw (43) to (27.center);
    \draw (26.center) to (44);
    \draw (44) to (45);
    \draw (45) to (28.center);
    \draw (42) to (44);
    \draw (43) to (45);
    \draw (43) to (44);

    \draw (64.center) to (51);
    \draw (66.center) to (54);
    \draw (51) to (54);
    \draw (65.center) to (58);
    \draw (67.center) to (63);
    \draw (58) to (63);
    \draw (54) to (61);
    \draw (52) to (58);
    \draw (54) to (58);

    \draw (70.center) to (76.center);
    \draw (84.center) to (83.center);
    \draw (79) to (75);
    \draw (79) to (73);
	\end{pgfonlayer}
\end{tikzpicture}
  %}
  \caption{Successive applications of rewriting rules to a simple ZX
  diagram (to be read from left to right and top to bottom).}
  \label{fig:running-example}
  %\end{center}
\end{figure}

This section introduces the basic rewriting rules of ZX calculus that
are outlined in
Figure~\ref{fig:rules}~\cite{coeckePicturingQuantumProcesses2017,staudacherReducing2QuBitGate2023}.
All rules remain valid under colour inversion.
We give an example of the application of successive rewriting rules,
explained below, on a simple ZX diagram in Figure~\ref{fig:running-example}.

\paragraph*{\textbf{Spider fusion (f)}}
Connected spiders of the same colour fuse through modulo-$2\pi$
addition of their phases. The reverse unfusing operation
is always possible, because connecting additional spiders with a
phase of $\alpha=0$ will not change the modulo-$2\pi$ addition. As a
consequence, infinite spiders can be unfused.
Figure~\ref{fig:running-example} highlights the fusion of two green non-phase-carrying spiders with their neighbouring phase-carrying spiders.

\paragraph*{\textbf{Local complementation (lc)}}
The local complementation rule~\cite{kotzigEulerianLinesFinite1968}
originates from graph theory. For all directly connected spiders of a
target spider, local complementation connects previously unconnected
spiders and disconnects previously connected spiders.
Local complementation of the highlighted red spider in the bottom
qubit row is illustrated in
Figure~\ref{fig:running-example}. The two green spiders connected to
the highlighted red spider connect via local complementation.
Performing a second local complementation at the same red spider
would disconnect the two green phase-carrying spiders again.
Pivoting describes a series of local complementations.

\paragraph*{\textbf{Colour change (h)}}
Adding Hadamard generators to each input and output inverts the colour of a spider. In Figure~\ref{fig:running-example},
all red spiders turn green with the addition of Hadamard generators.

\paragraph*{\textbf{Identity removal (i1, i2)}}
Non-phase-carrying spiders that are directly connected to other
spiders function as wires and leave the linear map of qubits unchanged.
The identity matrix $\underset{2^{m}\times 2^{n}}{\identity}$
represents the linear map of such spiders. A single wire replaces a
phaseless spider with $n=1$ and $m=1$. Similarly, two directly
connected Hadamard generators cancel each other out and act as a wire.
Applying the identity removal rule on the fused diagram in Figure
\ref{fig:running-example} removes all non-phase-carrying spiders that
possess one input and one output. Furthermore, identity rules are
used to convert a ZX diagram to be graph-like
(see Definition~\ref{def:graph-like}) by ensuring that spiders always connect with each other through Hadamard generators and by the addition of potentially
missing spiders at the input and output.

\paragraph*{\textbf{Bialgebra (b)}}
The bialgebra rule originates from the algebraic commutation relation
between the \textit{copy} and the \textit{or} gate. It allows connected
and opposite-coloured spiders to move through each other at the cost of
potentially adding spiders.

\paragraph*{\textbf{Copy ($\pi$, c)}}
$\pi$ copying moves an input spider that carries the phase
$\alpha=\pi$ through an opposite coloured spider to all connected
wires while multiplying the phase by $-1$.
If the input spider does not have any input wire ($n=0$) and the
phase is a multiple of $\pi$, the opposite coloured spider vanishes.
This second rule is referred to as the state copying, because it
copies the computational basis through an opposite-coloured spider.

\subsection{ZX-based Circuit Optimization}
\label{sec:zx_optimization}
Recent advances in quantum circuit optimization combine the ZX
calculus with algorithmic, heuristic or deep learning
approaches~\cite{duncanGraphtheoreticSimplificationQuantum2020,winderlRecursivelyPartitionedApproach2023,nagele}.
PyZX is a popular Python library to work with ZX calculus and
supports state-of-the-art circuit optimization with a focus on T-gate
reduction~\cite{kissingerPyZXLargeScale2020,kissingerReducingTcountZXcalculus2020}.
To simplify ZX diagrams using basic rewriting rules, PyZX assumes
that the ZX diagrams are graph-like.
\begin{definition}
  \label{def:graph-like}
  Graph-like ZX diagrams are only composed of Z spiders (green) that are connected by Hadamard wires. Input / Output possesses at most
  one wire that can only connect to one spider.
\end{definition}
The final diagram in Figure~\ref{fig:running-example} is graph-like. Every ZX
diagram can be converted to be graph-like using the h-rule and the
identity rules i1 and i2.

Reducing the number of T-gates in a given quantum circuit is crucial
because implementing the required quantum error correction on quantum
hardware demands significantly more resources compared to Clifford
gates~\cite{campbellRoadsFaulttolerantUniversal2017,dingMagicStateFunctionalUnits2018}.
% To reduce T-gate, PyZX relies on the full reduce algorithm, which
% is the current state-of-the-art algorithm.

The \textit{full reduce algorithm} is the main optimization algorithm of PyZX.
It aims to  decrease the number of T-gates of a given graph-like ZX
diagram by targeting spiders that carry phase a multiple of
$\frac{\pi}{2}$ and $\pi$~\cite{kissingerReducingTcountZXcalculus2020,duncanGraphtheoreticSimplificationQuantum2020}.
%Full reduce primarily aims to reduce the T-gate count by targeting
%interior, Clifford and Pauli spiders
%and phase gadgets.
%Interior spiders are connected with one input and one output.
%Clifford and Pauli spiders can act on the Z-basis (green) and X-basis
%(red) and are distinguished by the type of phase they carrying.
%Clifford spiders carry a phase that is a multiple of $\frac{\pi}{2}$
%and Pauli spiders carry a phase that is a multiple of $\pi$.
%Phase gadgets consists of a phase carrying spider with a single
%output that is connected with a Hadamard wire to a non-phase carrying spider.

After optimizing a ZX diagram, the corresponding quantum circuit
needs to be extracted. While converting a quantum circuit into ZX
diagram is a straightforward process, the opposite is
\#P-hard (at least as hard as NP-complete problems) and is known as
the circuit extraction
problem~\cite{debeaudrapCircuitExtractionZXdiagrams2022}. Polynomial time
algorithms exist for ZX diagrams that are graph-like and preserve the
generalized flow~\cite{backensThereBackAgain2020}. The disadvantage
of current circuit extraction algorithms is that the connectivity of
spiders in the ZX representation is replicated by two-qubit gates in
the resulting quantum circuit.

Recent advances optimize two-qubit gates by reducing the number of
edges in a given ZX diagram. Staudacher et al. proposed a heuristic
that calculates the cost based on the number of edges after rule
application and uses a greedy or stochastic algorithm to select
the next rule~\cite{staudacherReducing2QuBitGate2023}.

This paper focuses on ZX diagram optimization and not
on the circuit extraction problem. Therefore, the standard PyZX extraction algorithm is used for all
experiments~\cite{duncanGraphtheoreticSimplificationQuantum2020}.

\section{ZX Diagram Optimization}
\label{sec:dfs-for-zx}

Let $\mathbf{ZX}$ be the infinite set of all finite ZX diagrams,
$\mathbf{QC}$ the set of quantum circuits and $\mathbf{LM}$ the set
of linear maps of qubits.
We have two functions $\alpha: \mathbf{QC} \to \mathbf{ZX}$ and
$\mathit{extract}: \mathbf{ZX} \to \mathbf{QC} \cup \{\bot\}$ that
convert a quantum circuit into a ZX diagram and conversely.
The function $\mathit{extract}$ can map to a special element $\bot$
when it fails to extract a quantum circuit from a diagram.
Additionally, we have a function $\gamma: \mathbf{QC} \to
\mathbf{LM}$ that maps a quantum circuit to its linear map of qubits.

A \textit{quantum circuit optimization} algorithm is a function $f:
\mathbf{QC} \to \mathbf{QC}$ that optimizes some properties of the
quantum circuit.
We say that $f$ is \textit{semantic-preserving} whenever, for all $q
\in \mathbf{QC}$, we have $\gamma(q) = \gamma(f(q))$.

Let $R = \{h, b, lc, f, i1, i2, \pi, c, hd\}$ be the set of rewriting
rules presented in Section~\ref{sec:zxrules}.
A ZX rewriting rule $r \in R$ is a function $\mathbf{ZX} \to
\mathbf{ZX}$ such that the function $\mathit{extract} \circ r \circ
\alpha$ is semantic-preserving when the extraction succeeds.

A \textit{ZX-based quantum circuit optimization} algorithm is
searching for an extractable ZX diagram that optimizes one or more
properties of the quantum circuit.
Let $q \in \mathbf{QC}$ be a quantum circuit and $\mathit{opt}:
\mathbf{ZX} \to \mathbb{Z} \cup \{\bot\}$ be the optimization
function mapping a ZX diagram to a metric (e.g. the number of
T-gates, edges or two-qubit gates). The function $\mathit{opt}$ can
map the special element $\bot$ when it fails to compute a metric from
a ZX diagram.
Without loss of generality, we consider that we aim at minimizing
$\mathit{opt}$.
The \textit{ZX state-space} of $q$ is a set $W \subseteq \mathbf{ZX}$
such that $w \in W$ if there exists a finite sequence of rewriting
rules $r_1, \ldots, r_n$ such that $w = (r_n \circ \ldots \circ r_1 \circ
\alpha)(q)$.
The set of \textit{solutions} $S \subseteq W$ are all the extractable
ZX diagrams in $W$, that is, $\forall{w \in W},~\mathit{extract}(w)
\neq \bot \Leftrightarrow w \in S$.
The set of \textit{optimal solutions} is the largest set $O \subseteq
S$ such that for all $o \in O, s \in S$, we have
$\mathit{opt}(o) \leq \mathit{opt}(s)$.

There are challenges pertaining to the exploration of the ZX state-space.
Firstly, real-world quantum circuits result in large ZX diagrams with
a high memory demand for every state.
Secondly, the state-space to explore is infinite because the ZX rules
rewriting system is non-terminating, e.g. unfusing phaseless spiders
and colour changing is always possible.
Thirdly, to find a solution we must extract a circuit,
which is a computationally expensive operation that may fail if the
general flow is not preserved~\cite{backensThereBackAgain2020}.
Although it is possible to optimize a metric completely defined on
the ZX diagram (e.g. number of T-gates, vertices and edges), other metrics of
interest (e.g. number of two-qubits gates, circuit depth and overall
gate count) are defined on the extracted circuit.
Even though we focus in our experiments on the T-gate and edge count, our
approach is general and can be reused for any of those metrics.

In this paper, we rely on iterative deepening depth-first search
(IDDFS), which is a simple and efficient optimization algorithm, to
tackle these challenges~\cite{korfDepthfirstIterativedeepeningOptimal1985}.
More advanced optimization algorithms are difficult to use in the
context of ZX optimization.
For example, constraint-based combinatorial optimization such as
linear programming and constraint programming require  explicitly
encoding the rewriting rules as constraints and do not usually
support unbounded rewriting sequences out of the box.
Dynamic programming is an interesting approach to avoid re-exploring
the same state multiple times, but it requires to store a prohibitively
high number of states.
We would need to find a more compact representation of states, which
we leave for future work.

An interesting aspect of IDDFS is to strike the right balance between
exploration and exploitation.
It is based on depth-first search (DFS) which is a backtracking
algorithm applying rules in
sequence until it finds a leaf node, at which point it backtracks to
the previous decision made.
The problem with DFS is that wrong decisions taken early in the
search tree condemn the search strategy to explore large
uninteresting subtrees.
In the worst-case scenario, DFS reaches a deep leaf node with a state
that is unextractable.
Instead, IDDFS applies DFS with a depth bound, exploring successive
search trees of increasing depth.
It is especially useful when we do not have a good heuristic to
select the next node to explore, which is the case here, since it is
hard to predict which rules might lead to a better solution.
Another advantage of IDDFS is to use as much memory as DFS while
exploring the search tree in breadth as well.
Note that for completeness, we experiment with DFS when evaluating our approach.

A \textit{leaf node} is a ZX diagram $d \in \mathbf{ZX}$ such that
one of these conditions is true:
\begin{itemize}
  \item For all rules $r \in R\setminus\{h\}$ we have $r(d) = d$,
    that is, no rule besides colour change is able to rewrite the diagram.
  \item One of the pruning conditions is true.
\end{itemize}
When we reach a leaf node, we test whether $\mathit{opt}(d)$ is
better than the previously found solution, if any. In case of improvement, we check whether the
ZX diagram is extractable or not using $\mathit{extract}(d) = \bot$.
If it is extractable, we save the ZX diagram as the current best solution.

\textit{Pruning conditions} reduce the search tree effectively and
ensure that the search terminates in finite time.
%Rule-based pruning conditions prevent restrict the allowed sequence
% of rewriting rules.
%State-based pruning conditions terminate a path according to the
%properties of a given node, its ZX diagram or extracted quantum
%circuit.
The pruning conditions used throughout this paper are the following:
\begin{itemize}
  \item \textit{No spider unfusion.} It is always possible to
    separate one spider into multiple spiders as long as the
    $\text{mod } \pi$ sum of all involved spiders remains unchanged.
  \item \textit{Rule bundling.} If a rule can rewrite a given ZX
    diagram more than once, all possible modifications are performed
    in a batch, hence generating only one node in the search tree.
  \item \textit{No colour cycle.} Disallowing consecutive colour
    changing rule application avoids infinite paths that consist only
    of recolouring spiders.
  \item \textit{Global time limit.} After a set time limit is
    exceeded, the search is terminated and the best solution found so
    far is returned.
\end{itemize}

\section{Computational Experiments}

The proposed DFS and IDDFS approaches are implemented in Python and
evaluated against PyZX's implementation of full reduce using a
diverse set of quantum
circuits~\cite{duncanGraphtheoreticSimplificationQuantum2020,namGitHubMeamyTpar2019}.

Although the search algorithms employed are fairly simple and
well-known, their combination with ZX-calculus and the pruning
condition implemented in a reproducible and integrated framework is not
straightforward. A tight integration with PyZX and Qiskit, as well as
the contribution
of a Qiskit transpiler pass, ensures the reusability of our approach.
The overall project is 7000 lines of code.

Every search instance was executed on a Xeon Gold 6132
clocked at 2.6 GHz and 64 GB of 2400 MHz DDR4 of available RAM running
Rocky Linux 8.10 with Python 3.12.

We evaluated the performance of the various algorithms on the complete
\textit{set of 100 standard quantum circuits} using the pruning
conditions introduced in Section~\ref{sec:dfs-for-zx}. A
\textit{global timeout of 1.5 hours} is set for every instance. The
rules are ordered such that a change connectivity (e.g.
pivoting and local complementation) takes priority over spider count
reduction (e.g. fusion and identity removal). Upon completion of an instance, the optimized quantum circuit can be fed into a transpilation pipeline for further processing.
In a first set of experiments, we minimized the T-gate count of ZX
diagrams. To demonstrate the generality of our approach, a second set
of experiments that minimizes the edge count of ZX diagrams is
executed. The full results for every instance, including a comparison with full-reduce and the algorithm runtimes, are available in the supplementary material.

\subsection{T-gate Reduction}
\label{sec:t-gate-reduction}

\begin{figure*}[ht!]
  \fontfamily{DejaVuSans-TLF}
  \selectfont
  \centering
  Solution comparison of IDDFS and DFS\par\medskip
  \begin{subfigure}{0.5\textwidth}
    \includegraphics[width=\textwidth,keepaspectratio]{./analysis/time\_instances\_solved.eps}
    \caption{Time evolution against the solution of full reduce.}
    \label{fig:time-instances-solved}
  \end{subfigure}%
  \begin{subfigure}{0.5\textwidth}
    \includegraphics[width=\textwidth,keepaspectratio]{./analysis/time\_instances\_solved\_edges.eps}
    \caption{Time evolution of the best solution found.}
    \label{fig:time-instances-edge-count}
  \end{subfigure}
\end{figure*}

The DFS search exhibits poorer performance, only equating full
reduce in $46\%$ of the instances, compared to IDDFS search. IDDFS
equates full reduce in $89\%$ of the instances. Neither
DFS nor IDDFS approaches are able to outperform full reduce within
the 1.5 hour time limit. Nevertheless, it should be noted that the DFS
search is able to equate full reduce on three circuits on which IDDFS
leads to poorer results.

Figure~\ref{fig:time-instances-solved} shows the time evolution of
the best solution of DFS (red) and IDDFS (blue). As neither DFS nor
IDDFS outperform full reduce, only solutions that equate full reduce
are shown. DFS almost immediately equals full reduce in
$41\%$ of the instances and only equating full reduce on additional
$5\%$ of the instances for the remaining 1.5 hours. In contrast,
IDDFS requires $16$ minutes to level the performance of DFS and equals
full reduce in $80\%$ of the instances within the first $60$ minutes.
Overall, IDDFS equals full reduce on $89\%$ of the instances within
the 1.5 hour time limit.

\subsection{Edge Reduction}
Figure~\ref{fig:time-instances-edge-count} visualizes the time
evolution of the best solution of DFS (red) and IDDFS (blue). Within
the 1.5 hour time limit, IDDFS is able to find the best solution in
$86\%$ of the instances.

DFS results only in the best solution exclusively in $1\%$ of the
instances and shares the best solution in $31\%$ of the instances.
IDDFS improves the performance of DFS and leads to an exclusively
best solution for $54\%$ of the instances and shares its best
solution in $32\%$ of the instances.

It should be noted, that despite being designed to reduce the T-gate
count, full reduce achieves the exclusively best solution on $13\%$ of
the instances and shares a best solution in an additional $2\%$ of
the instances.

Compared to the unoptimized quantum circuit, DFS
improves the edge count on average by $11\%$. IDDFS exhibits better
performance and reduces the edge count by $22\%$ within the 1.5 hour
time limit. Full reduce improves the edge count by $3\%$.
We optimized the edge count to demonstrate the applicability of our
algorithm to other metrics. Furthermore, optimizing the edge count is
interesting because current state-of-the-art circuit extraction algorithms replicate the connectivity of spiders with two-qubit gates, therefore potentially increasing the circuit depth and two-qubit gate count.

Our results are more contrasted as reducing the edge count did not
necessarily translate in a reduction of the of two-qubit gate count.
On the contrary, more two-qubit gates are added for the vast majority
of instances. DFS results in a higher two-qubit gate count in
$41\%$ of the instances and only reduces the two-qubit gate count in
$1\%$ of the instances. IDDFS yields to poorer performance and adds
two-qubit gates in $85\%$ of the instances and reduces the two-qubit
gate count in $1\%$ of the instances.
translates in a reduction of the two-qubit gate count. Staudacher
et al. showed that they were able to translate an average reduction in the
edge count by $29\%$ in a reduction of the two-qubit gate count by
$21\%$~\cite{staudacherReducing2QuBitGate2023}.

\section{Related Work}
\label{sec:related-work}

Recent advances have been made to optimize the T-gate count in
quantum circuits. Fault tolerant quantum computing introduces quantum
error correction code that increases the resource demand, especially
for the T-gate. Improvements in the T-gate count that leverages the
quantum error code were achieved with Matroid
partitioning~\cite{amyPolynomialTimeTDepthOptimization2014}.
Template based techniques improve the quantum circuit synthesis by
reducing the T-gate count and circuit
depth~\cite{biswalTemplatebasedTechniqueEfficient2018}.

% ZX calculus minimizaiton
The first proposed optimization strategy using ZX calculus is
restricted to Clifford
gates~\cite{faganOptimisingCliffordCircuits2019}. The
state-of-the-art optimization algorithm full reduce targets Clifford
and
T-gates~\cite{duncanGraphtheoreticSimplificationQuantum2020,kissingerReducingTcountZXcalculus2020}.
Other techniques optimize the
T-gate count through the treatment of Clifford gates
and Pauli operators as $\frac{\pi}{4}$ rotations around each
other~\cite{zhangOptimizingGatesClifford+T2019}.
Additionally, new causal flow preserving optimization techniques
ensure the extractability of a quantum circuit from a ZX
diagram~\cite{holkerCausalFlowPreserving2024}. An improved T-gate count
for arithmetic circuits, e.g. integer multiplication, was found
by applying the ZX rewriting
rules~\cite{joshiQuantumCircuitOptimization2023}.
Reinforcement learning strategies based on ZX calculus that target
the T-gate and two-qubit gate count emerged in recent
years~\cite{nagele,riuReinforcementLearningBased2023}.
Heuristics that target the two-qubit counts ensures the usefulness of
ZX calculus for photonic quantum computing and other quantum hardware
that does not perform error
correction~\cite{staudacherReducing2QuBitGate2023}. Other approaches
combine heuristics and ZX calculus for the architecture-aware
optimization of quantum
circuits~\cite{gogiosoAnnealingOptimisationMixed2023,winderlRecursivelyPartitionedApproach2023}.

Heuristic approaches deal with the time complexity involved in
quantum circuit optimization. In principle, a heuristic
pattern matching algorithm is combined with gate commutation rules to
minimize the total gate
count~\cite{itenExactPracticalPattern2022}.
Boolean satisfiability is an exact approach for the optimization of
classical circuits. Despite the challenging encoding of quantum
gates, advances have been made to bring this approach to quantum
circuit
optimization~\cite{meuliSATbasedCNOTQuantum2018,berentSATEncodingQuantum2022}.
Recently, reinforcement learning techniques emerged for quantum
circuit optimization and mapping of quantum circuits for specific quantum
architectures~\cite{foselQuantumCircuitOptimization2021,elsayedamerOptimalityQuantumCircuit2024}.
Gate commutation rules and templates proved also advantageous for the
mapping of quantum
circuits~\cite{itokoOptimizationQuantumCircuit2020}.

\section{Conclusion}
\label{sec:conclusion}
This paper lays the foundation to apply exhaustive search to ZX
diagrams for the optimization of quantum circuits. The combination of
the semantics-preserving rewriting rule of ZX calculus with the exhaustive
search algorithms \textit{depth-first search (DFS)} and
\textit{iterative deepening depth-first search (IDDFS)} enables to
target metrics of the ZX diagram or its corresponding quantum circuit
without being designed for one specific metric.

Our results indicate that IDDFS is a more effective approach for ZX
diagram optimization than DFS. Within the 1.5 hour time limit IDDFS
is able to equate state-of-the-art algorithms that reduce the T-gate
count in $89\%$ of the instances and competes with novel approaches that
reduce the edge count, demonstrating the applicability of our approach.

A Qiskit compiler pass that implements the DFS and IDDFS approach,
with configurable pruning conditions and integration with the PyZX
library, is available on GitLab (\url{https://gitlab.com/NetForceExplorer/zx_dfs/-/releases/v1.0-OLA_2025}).

Our results demonstrate that not every reduction in the edge count
translates into a reduction in the two-qubit gate count. Upcoming work
could focus on the enhancement of the edge count metric to better
approximate the two-qubit gate count after circuit extraction.
Future efforts should address the scalability issue for large
circuits of the IDDFS and DFS based optimization. The principal idea
of ZX diagram optimization is to change the connectivity and the
fusion of spiders, hence a \textit{limited discrepancy search} could improve the performance~\cite{Harvey1995}. The application of
dynamic programming techniques could trade computational performance for higher memory requirements.
Finally, bridging the gap from architecture-independent
optimization towards quantum architecture-aware optimization could
address the execution of real-world quantum circuits on
next-generation hardware.

\begin{credits}
  Tobias Fischbach acknowledges financial support from the
Institute for Advanced Studies of
  the University of Luxembourg through a YOUNG ACADEMICS Grant
(YOUNG ACADEMICS-
  2022-NETCOM)
\end{credits}

\bibliographystyle{splncs04}
\bibliography{bibliography}

\section*{Supplementary Material}
\subsection*{T-Gate Count}
\begin{figure}[h!]
  \begin{center}
    \includegraphics{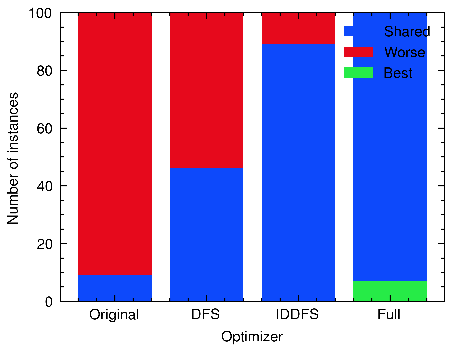}
    \caption{Performance comparison of the best solution found.}
  \end{center}
\end{figure}

\begin{longtable}{lrrrrr}
\hline
 Circuit              & Original   & DFS                & IDDFS              & Full reduce         \\
\hline
 2of5d1               & 167        & 167\quad 00:00:00      & \textBF{87\quad 00:06:54}   & \textBF{87\quad 00:00:00}    \\
 2of5d1s              & 129        & 129\quad 00:00:00      & \textBF{79\quad 00:05:23}   & \textBF{79\quad 00:00:02}    \\
 2of5d2               & 43         & 37\quad 00:00:00       & \textBF{35\quad 00:08:12}   & \textBF{35\quad 00:00:03}    \\
 2of5d3               & 102        & 96\quad 00:00:00       & \textBF{52\quad 00:08:33}   & \textBF{52\quad 00:00:03}    \\
 3-17tc               & \textBF{12}     & \textBF{12\quad 00:00:00}   & \textBF{12\quad 00:00:00}   & \textBF{12\quad 00:00:00}    \\
 4-49-12-32           & 33         & \textBF{29\quad 00:00:00}   & \textBF{29\quad 00:39:35}   & \textBF{29\quad 00:00:00}    \\
 4-49tc1              & 67         & \textBF{49\quad 00:00:01}   & \textBF{49\quad 00:31:46}   & \textBF{49\quad 00:00:00}    \\
 4b15g-1              & 50         & \textBF{42\quad 00:00:00}   & \textBF{42\quad 00:46:28}   & \textBF{42\quad 00:00:01}    \\
 4b15g-2              & 74         & \textBF{62\quad 00:00:00}   & \textBF{62\quad 00:22:59}   & \textBF{62\quad 00:00:01}    \\
 4b15g-3              & 64         & \textBF{50\quad 00:00:00}   & \textBF{50\quad 00:25:05}   & \textBF{50\quad 00:00:02}    \\
 4b15g-4              & 52         & \textBF{44\quad 00:00:00}   & \textBF{44\quad 00:35:16}   & \textBF{44\quad 00:00:04}    \\
 4b15g-5              & 45         & \textBF{37\quad 00:00:00}   & \textBF{37\quad 00:17:12}   & \textBF{37\quad 00:00:05}    \\
 5bit-adder            & 55         & 55\quad 00:00:00       & \textBF{39\quad 00:18:12}   & \textBF{39\quad 00:00:00}    \\
 5mod5-10-71a         & 55         & 55\quad 00:00:00       & \textBF{43\quad 00:15:07}   & \textBF{43\quad 00:00:00}    \\
 5mod5-8              & 67         & 67\quad 00:00:00       & \textBF{43\quad 00:17:29}   & \textBF{43\quad 00:00:01}    \\
 5mod5-fc             & 72         & 72\quad 00:00:00       & \textBF{44\quad 00:22:09}   & \textBF{44\quad 00:00:02}    \\
 5mod5tc              & 178        & 178\quad 00:00:00      & \textBF{76\quad 00:20:25}   & \textBF{76\quad 00:00:00}    \\
 6symd2               & 81         & 73\quad 00:00:00       & \textBF{67\quad 00:15:23}   & \textBF{67\quad 00:00:01}    \\
 9symd2               & 126        & 114\quad 00:00:00      & \textBF{108\quad 00:07:39}  & \textBF{108\quad 00:00:02}   \\
 gf$2^{10}$-mult-109-509   & 600        & 600\quad 00:00:00      & \textBF{410\quad 00:29:30}  & \textBF{410\quad 00:00:08}   \\
 gf$2^{11}$-mult-131-615   & 737        & 737\quad 00:00:00      & \textBF{517\quad 00:29:59}  & \textBF{517\quad 00:00:08}   \\
 gf$2^{12}$-mult-177-753   & 864        & 864\quad 00:00:00      & \textBF{588\quad 00:26:05}  & \textBF{588\quad 00:00:39}   \\
 gf$2^{13}$-mult-205-881   & 1027       & 1027\quad 00:00:00     & \textBF{715\quad 00:31:55}  & \textBF{715\quad 00:01:31}   \\
 gf$2^{14}$-mult-235-1019  & 1176       & 1176\quad 00:00:00     & \textBF{798\quad 00:24:14}  & \textBF{798\quad 00:02:27}   \\
 gf$2^{15}$-mult-239-1139  & 1365       & 1365\quad 00:00:00     & \textBF{945\quad 00:31:58}  & \textBF{945\quad 00:00:21}   \\
 gf$2^{16}$-mult-301-1325  & 1536       & 1536\quad 00:00:00     & \textBF{1040\quad 00:48:57} & \textBF{1040\quad 00:01:38}  \\
 gf$2^{17}$-mult-305-1461  & 1751       & 1751\quad 00:00:00     & \textBF{1207\quad 00:43:28} & \textBF{1207\quad 00:03:24}  \\
 gf$2^{18}$-mult-375-1671  & 1944       & 1944\quad 00:00:00     & 1324\quad 00:36:05     & \textBF{1314\quad 00:05:23}  \\
 gf$2^{19}$-mult-415-1859  & 2185       & 2185\quad 00:00:00     & \textBF{1501\quad 00:52:45} & \textBF{1501\quad 00:00:55}  \\
 gf$2^{20}$-mult-419-2019  & 2400       & 2400\quad 00:00:00     & 1632\quad 01:06:32     & \textBF{1620\quad 00:04:11}  \\
 gf$2^{32}$-mult-1117-5213 & 6144       & 6144\quad 00:00:00     & 6144\quad 00:00:00     & \textBF{4128\quad 00:14:33}  \\
 gf$2^{32}$-mult-1148-5244 & 6144       & 6144\quad 00:00:00     & 6144\quad 00:00:00     & \textBF{4128\quad 00:41:38}  \\
 gf$2^{32}$-mult-1179-5275 & 6144       & 6144\quad 00:00:00     & 6144\quad 00:00:00     & \textBF{4128\quad 00:08:39}  \\
 gf$2^{3}$-mult-11-47      & 57         & 57\quad 00:00:00       & \textBF{45\quad 00:18:29}   & \textBF{45\quad 00:31:21}    \\
 gf$2^{4}$-mult-19-83      & 96         & 96\quad 00:00:00       & \textBF{68\quad 00:14:25}   & \textBF{68\quad 00:31:22}    \\
 gf$2^{50}$-2647-12647    & 15000      & 15000\quad 00:00:00    & 15000\quad 00:00:00    & \textBF{10050\quad 01:29:21} \\
 gf$2^{5}$-mult-29-129     & 155        & 155\quad 00:00:00      & \textBF{115\quad 00:20:00}  & \textBF{115\quad 00:00:00}   \\
 gf$2^{6}$-mult-41-185     & 216        & 216\quad 00:00:00      & \textBF{150\quad 00:13:42}  & \textBF{150\quad 08:36:33}   \\
 gf$2^{7}$-mult-55-251     & 301        & 301\quad 00:00:00      & \textBF{217\quad 00:15:56}  & \textBF{217\quad 08:36:36}   \\
 gf$2^{8}$-mult-85-341     & 384        & 384\quad 00:00:00      & \textBF{264\quad 00:08:33}  & \textBF{264\quad 00:00:02}   \\
 gf$2^{9}$-mult-89-413     & 495        & 495\quad 00:00:00      & \textBF{351\quad 00:11:39}  & \textBF{351\quad 00:00:11}   \\
 graycode6            & \textBF{0}      & \textBF{0\quad 00:00:00}    & \textBF{0\quad 00:00:00}    & \textBF{0\quad 00:00:18}     \\
 ham15-109-214        & 141        & \textBF{99\quad 00:00:07}   & \textBF{99\quad 00:18:21}   & \textBF{99\quad 00:00:19}    \\
 ham15-70             & 498        & \textBF{192\quad 00:00:24}  & \textBF{192\quad 00:29:39}  & \textBF{192\quad 00:00:03}   \\
 ham3tc               & \textBF{7}      & \textBF{7\quad 00:00:00}    & \textBF{7\quad 00:00:00}    & \textBF{7\quad 00:00:10}     \\
 ham7-21-69           & 67         & \textBF{45\quad 00:00:00}   & \textBF{45\quad 00:27:55}   & \textBF{45\quad 00:00:10}    \\
 ham7-25-49           & 42         & \textBF{30\quad 00:00:00}   & \textBF{30\quad 01:10:43}   & \textBF{30\quad 00:00:11}    \\
 ham7tc               & 91         & \textBF{59\quad 00:00:00}   & \textBF{59\quad 00:15:02}   & \textBF{59\quad 00:00:01}    \\
 hwb4-11-21           & \textBF{21}     & \textBF{21\quad 00:00:00}   & \textBF{21\quad 00:00:00}   & \textBF{21\quad 00:00:03}    \\
 hwb4-11-23           & \textBF{21}     & \textBF{21\quad 00:00:00}   & \textBF{21\quad 00:00:00}   & \textBF{21\quad 00:00:04}    \\
 hwb4tc               & 72         & \textBF{50\quad 00:00:00}   & \textBF{50\quad 00:23:16}   & \textBF{50\quad 00:00:05}    \\
 hwb5-24-102          & 124        & \textBF{78\quad 00:00:01}   & \textBF{78\quad 00:20:47}   & \textBF{78\quad 00:00:01}    \\
 hwb5-24-114          & 124        & \textBF{78\quad 00:00:01}   & \textBF{78\quad 00:23:47}   & \textBF{78\quad 00:00:05}    \\
 hwb5-31-91           & 95         & \textBF{73\quad 00:00:00}   & \textBF{73\quad 00:16:11}   & \textBF{73\quad 00:00:08}    \\
 hwb5tc               & 369        & \textBF{239\quad 00:00:17}  & \textBF{239\quad 00:22:05}  & \textBF{239\quad 00:00:15}   \\
 hwb6-42-150          & 155        & \textBF{125\quad 00:00:03}  & \textBF{125\quad 00:17:16}  & \textBF{125\quad 00:00:02}   \\
 hwb6-47-107          & 97         & \textBF{75\quad 00:00:01}   & \textBF{75\quad 00:23:21}   & \textBF{75\quad 00:00:06}    \\
 hwb6tc               & 1709       & \textBF{911\quad 00:02:13}  & 923\quad 01:23:58      & \textBF{911\quad 00:00:53}   \\
 hwb7-236             & 4252       & \textBF{2178\quad 00:10:32} & 2256\quad 01:07:05     & \textBF{2178\quad 00:05:39}  \\
 hwb7-331-2609a       & 2461       & \textBF{1397\quad 00:06:47} & 1403\quad 00:28:55     & \textBF{1397\quad 00:00:29}  \\
 hwb7tc               & 5396       & \textBF{2596\quad 00:19:11} & 3472\quad 00:09:48     & \textBF{2596\quad 00:02:43}  \\
 hwb8-2710-6940       & 5405       & 5405\quad 00:00:00     & 4247\quad 01:25:31     & \textBF{3503\quad 00:06:31}  \\
 mod5-adder-15         & 107        & \textBF{75\quad 00:00:00}   & \textBF{75\quad 00:07:17}   & \textBF{75\quad 00:09:36}    \\
 mod5-adder-17-81      & 95         & \textBF{71\quad 00:00:00}   & \textBF{71\quad 00:24:33}   & \textBF{71\quad 00:00:00}    \\
 mod5-adders           & 136        & 136\quad 00:00:00      & \textBF{86\quad 00:15:06}   & \textBF{86\quad 00:00:01}    \\
 mod5d1               & 22         & 22\quad 00:00:00       & \textBF{8\quad 00:17:43}    & \textBF{8\quad 00:00:01}     \\
 mod5d2               & 22         & 22\quad 00:00:00       & \textBF{8\quad 00:07:07}    & \textBF{8\quad 00:00:02}     \\
 mod5d4               & \textBF{7}      & \textBF{7\quad 00:00:00}    & \textBF{7\quad 00:00:00}    & \textBF{7\quad 00:00:00}     \\
 mod5mils             & 12         & 12\quad 00:00:00       & \textBF{8\quad 00:24:29}    & \textBF{8\quad 00:00:00}     \\
 mspk-4-49-12         & 33         & \textBF{29\quad 00:00:00}   & \textBF{29\quad 00:45:52}   & \textBF{29\quad 00:00:00}    \\
 mspk-4-49-13         & 33         & \textBF{29\quad 00:00:00}   & \textBF{29\quad 00:53:32}   & \textBF{29\quad 00:00:00}    \\
 mspk-4-49-14         & 31         & \textBF{29\quad 00:00:00}   & \textBF{29\quad 00:31:00}   & \textBF{29\quad 00:00:00}    \\
 mspk-4b15g-1         & 45         & \textBF{41\quad 00:00:00}   & \textBF{41\quad 01:07:05}   & \textBF{41\quad 00:00:00}    \\
 mspk-4b15g-2         & \textBF{33}     & \textBF{33\quad 00:00:00}   & \textBF{33\quad 00:00:00}   & \textBF{33\quad 00:00:00}    \\
 mspk-4b15g-3         & 33         & \textBF{29\quad 00:00:00}   & \textBF{29\quad 00:38:37}   & \textBF{29\quad 00:00:00}    \\
 mspk-4b15g-4         & 38         & \textBF{34\quad 00:00:00}   & \textBF{34\quad 00:49:26}   & \textBF{34\quad 00:00:00}    \\
 mspk-4b15g-5         & 38         & \textBF{34\quad 00:00:00}   & \textBF{34\quad 00:32:05}   & \textBF{34\quad 00:00:01}    \\
 mspk-hwb4-12         & \textBF{21}     & \textBF{21\quad 00:00:00}   & \textBF{21\quad 00:00:00}   & \textBF{21\quad 00:00:02}    \\
 mspk-hwb4-13         & \textBF{21}     & \textBF{21\quad 00:00:00}   & \textBF{21\quad 00:00:00}   & \textBF{21\quad 00:00:03}    \\
 mspk-nth-primes4-11  & 65         & \textBF{51\quad 00:00:01}   & \textBF{51\quad 00:27:18}   & \textBF{51\quad 00:00:00}    \\
 mspk-nth-primes4-12  & 40         & \textBF{38\quad 00:00:00}   & \textBF{38\quad 00:09:02}   & \textBF{38\quad 00:00:00}    \\
 mspk-nth-primes4-13  & 38         & \textBF{30\quad 00:00:00}   & \textBF{30\quad 00:25:58}   & \textBF{30\quad 00:00:00}    \\
 mspk-nth-primes4-14  & 35         & \textBF{31\quad 00:00:00}   & \textBF{31\quad 00:06:54}   & \textBF{31\quad 00:00:01}    \\
 or5d1                & 43         & 43\quad 00:00:00       & \textBF{31\quad 00:10:01}   & \textBF{31\quad 00:00:00}    \\
 or5d2                & 62         & \textBF{56\quad 00:00:00}   & \textBF{56\quad 00:08:44}   & \textBF{56\quad 00:00:02}    \\
 rd53-16-67           & 69         & 69\quad 00:00:00       & \textBF{43\quad 00:09:23}   & \textBF{43\quad 00:00:04}    \\
 rd53d15              & 121        & 119\quad 00:00:00      & \textBF{67\quad 00:13:53}   & \textBF{67\quad 00:00:07}    \\
 rd53d15s             & 115        & 115\quad 00:00:00      & \textBF{67\quad 00:11:21}   & \textBF{67\quad 00:00:01}    \\
 rd53d1               & 152        & 150\quad 00:00:00      & \textBF{60\quad 00:07:52}   & \textBF{60\quad 00:00:00}    \\
 rd53d1mils           & 86         & 86\quad 00:00:00       & \textBF{58\quad 00:06:29}   & \textBF{58\quad 00:00:02}    \\
 rd53d2               & 50         & 44\quad 00:00:00       & \textBF{38\quad 00:09:27}   & \textBF{38\quad 00:00:03}    \\
 rd53rcmg             & 269        & 269\quad 00:00:00      & \textBF{147\quad 00:13:06}  & \textBF{147\quad 00:00:01}   \\
 rd73d2               & 88         & 78\quad 00:00:00       & \textBF{70\quad 00:18:27}   & \textBF{70\quad 00:00:03}    \\
 rd84d1               & 129        & 117\quad 00:00:01      & \textBF{109\quad 00:09:15}  & \textBF{109\quad 00:00:04}   \\
 t6 1 52              & 74         & 74\quad 00:00:00       & \textBF{50\quad 00:15:12}   & \textBF{50\quad 00:00:06}    \\
 t6 3 48              & 72         & 72\quad 00:00:00       & \textBF{40\quad 00:14:52}   & \textBF{40\quad 00:00:00}    \\
 t7 1 84              & 100        & 100\quad 00:00:00      & \textBF{58\quad 00:15:03}   & \textBF{58\quad 00:00:01}    \\
 t7 4 64              & 96         & 96\quad 00:00:00       & \textBF{52\quad 00:20:33}   & \textBF{52\quad 00:00:02}    \\
 t8 1 116             & 124        & 124\quad 00:00:00      & \textBF{78\quad 00:20:16}   & \textBF{78\quad 00:00:03}    \\
 t8 5 80              & 120        & 120\quad 00:00:00      & \textBF{64\quad 00:06:30}   & \textBF{64\quad 00:00:00}    \\
\hline
\end{longtable}

\subsection*{Edge Count}

\begin{figure}[h!]
  \begin{center}
    \includegraphics{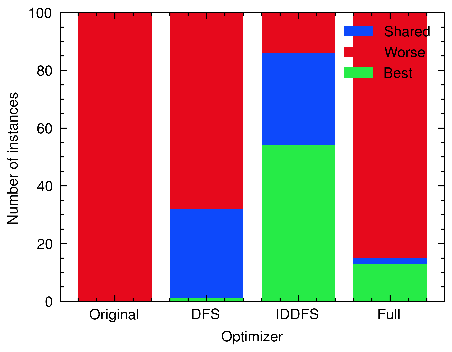}
    \caption{Performance comparison of the best solution found.}
  \end{center}
\end{figure}
\begin{longtable}{lrrrrr}
\hline
 Circuit              &   Original & DFS                 & IDDFS               & Full reduce       \\
\hline
 2of5d1               &        551 & 515\quad 00:00:01       & \textBF{334\quad 00:11:14}   & 465\quad 00:00:00     \\
 2of5d1s              &        428 & 394\quad 00:00:00       & \textBF{302\quad 00:11:25}   & 342\quad 00:00:02     \\
 2of5d2               &        149 & 134\quad 00:00:00       & \textBF{128\quad 00:26:09}   & 175\quad 00:00:03     \\
 2of5d3               &        341 & 311\quad 00:00:00       & \textBF{205\quad 00:12:37}   & 215\quad 00:00:03     \\
 3-17tc               &         50 & \textBF{41\quad 00:00:29}    & \textBF{41\quad 00:00:46}    & 50\quad 00:00:00      \\
 4-49-12-32           &        110 & \textBF{104\quad 00:00:00}   & \textBF{104\quad 00:39:35}   & 115\quad 00:00:00     \\
 4-49tc1              &        236 & 213\quad 01:12:23       & \textBF{192\quad 01:17:53}   & 227\quad 00:00:00     \\
 4b15g-1              &        177 & 156\quad 00:56:23       & \textBF{153\quad 00:58:01}   & 174\quad 00:00:01     \\
 4b15g-2              &        240 & 215\quad 00:00:00       & \textBF{214\quad 00:40:59}   & 329\quad 00:00:01     \\
 4b15g-3              &        215 & 184\quad 00:04:10       & \textBF{179\quad 00:57:07}   & 213\quad 00:00:02     \\
 4b15g-4              &        176 & \textBF{145\quad 00:46:26}   & 146\quad 00:48:00       & 171\quad 00:00:04     \\
 4b15g-5              &        156 & \textBF{130\quad 00:18:48}   & \textBF{130\quad 00:22:16}   & 140\quad 00:00:05     \\
 5bit-adder            &        240 & 220\quad 00:00:00       & \textBF{176\quad 00:34:18}   & 191\quad 00:00:00     \\
 5mod5-10-71a         &        201 & 187\quad 00:00:00       & 167\quad 00:42:36       & \textBF{156\quad 00:00:00} \\
 5mod5-8              &        238 & 219\quad 00:00:00       & \textBF{175\quad 00:28:34}   & 186\quad 00:00:01     \\
 5mod5-fc             &        261 & 244\quad 00:00:00       & 177\quad 00:27:40       & \textBF{172\quad 00:00:02} \\
 5mod5tc              &        610 & 562\quad 00:00:00       & 308\quad 00:25:56       & \textBF{294\quad 00:00:00} \\
 6symd2               &        264 & \textBF{245\quad 00:00:00}   & \textBF{245\quad 00:15:21}   & 323\quad 00:00:01     \\
 9symd2               &        398 & 363\quad 00:00:00       & \textBF{352\quad 00:25:54}   & 485\quad 00:00:02     \\
 gf$2^{10}$-mult-109-509   &       2077 & 1886\quad 00:00:02      & \textBF{1466\quad 01:12:09}  & 1581\quad 00:00:08    \\
 gf$2^{11}$-mult-131-615   &       2502 & 2275\quad 00:00:04      & \textBF{1749\quad 00:55:05}  & 1959\quad 00:00:08    \\
 gf$2^{12}$-mult-177-753   &       2999 & 2722\quad 00:00:21      & \textBF{2097\quad 00:38:58}  & 3270\quad 00:00:39    \\
 gf$2^{13}$-mult-205-881   &       3504 & 3183\quad 00:00:08      & \textBF{2464\quad 00:51:09}  & 3776\quad 00:01:31    \\
 gf$2^{14}$-mult-235-1019  &       4055 & 3680\quad 00:00:11      & \textBF{2827\quad 00:38:27}  & 5022\quad 00:02:27    \\
 gf$2^{15}$-mult-239-1139  &       4608 & 4185\quad 00:00:48      & \textBF{3175\quad 00:45:25}  & 3842\quad 00:00:21    \\
 gf$2^{16}$-mult-301-1325  &       5273 & 4780\quad 00:03:52      & \textBF{3651\quad 01:07:04}  & 5735\quad 00:01:38    \\
 gf$2^{17}$-mult-305-1461  &       5906 & 5351\quad 00:01:07      & \textBF{4023\quad 00:58:45}  & 5263\quad 00:03:24    \\
 gf$2^{18}$-mult-375-1671  &       6651 & 6028\quad 00:01:37      & \textBF{4571\quad 00:44:43}  & 8847\quad 00:05:23    \\
 gf$2^{19}$-mult-415-1859  &       7400 & 6705\quad 00:01:58      & \textBF{5060\quad 00:50:54}  & 8697\quad 00:00:55    \\
 gf$2^{20}$-mult-419-2019  &       8147 & 7376\quad 00:09:43      & \textBF{5519\quad 01:13:34}  & 6824\quad 00:04:11    \\
 gf$2^{32}$-mult-1117-5213 &      20785 & \textBF{18780\quad 01:24:51} & \textBF{18780\quad 01:21:55} & 23889\quad 00:14:33   \\
 gf$2^{32}$-mult-1148-5244 &      20812 & \textBF{18813\quad 01:20:04} & \textBF{18813\quad 01:18:24} & 26270\quad 00:41:38   \\
 gf$2^{32}$-mult-1179-5275 &      20843 & \textBF{18844\quad 00:26:01} & \textBF{18844\quad 00:28:02} & 29108\quad 00:08:39   \\
 gf$2^{3}$-mult-11-47      &        204 & 189\quad 00:00:00       & \textBF{163\quad 00:43:42}   & 174\quad 00:31:21     \\
 gf$2^{4}$-mult-19-83      &        351 & 324\quad 00:00:00       & \textBF{276\quad 00:33:04}   & 281\quad 00:31:22     \\
 gf$2^{50}$-2647-12647    &      50457 & \textBF{45546\quad 00:27:15} & \textBF{45546\quad 00:22:51} & 64331\quad 01:29:21   \\
 gf$2^{5}$-mult-29-129     &        540 & 493\quad 00:00:01       & \textBF{401\quad 00:49:11}   & 430\quad 00:00:00     \\
 gf$2^{6}$-mult-41-185     &        765 & 702\quad 00:00:02       & \textBF{565\quad 00:39:17}   & 594\quad 08:36:33     \\
 gf$2^{7}$-mult-55-251     &       1032 & 945\quad 00:00:03       & \textBF{751\quad 00:35:02}   & 894\quad 08:36:36     \\
 gf$2^{8}$-mult-85-341     &       1359 & 1236\quad 00:00:14      & \textBF{990\quad 00:18:54}   & 1425\quad 00:00:02    \\
 gf$2^{9}$-mult-89-413     &       1692 & 1533\quad 00:00:07      & \textBF{1181\quad 00:24:47}  & 1448\quad 00:00:11    \\
 graycode6            &         25 & 25\quad 00:00:00        & 25\quad 00:19:40        & \textBF{21\quad 00:00:18}  \\
 ham15-109-214        &        651 & 617\quad 00:00:02       & \textBF{584\quad 00:41:51}   & 625\quad 00:00:19     \\
 ham15-70             &       1722 & 1273\quad 00:30:22      & \textBF{997\quad 00:31:46}   & 1740\quad 00:00:03    \\
 ham3tc               &         34 & \textBF{24\quad 00:00:00}    & \textBF{24\quad 00:00:00}    & \textBF{24\quad 00:00:10}  \\
 ham7-21-69           &        249 & \textBF{188\quad 01:26:26}   & \textBF{188\quad 00:29:02}   & 240\quad 00:00:10     \\
 ham7-25-49           &        166 & \textBF{141\quad 00:03:11}   & \textBF{141\quad 01:13:55}   & 155\quad 00:00:11     \\
 ham7tc               &        334 & 242\quad 01:00:00       & \textBF{226\quad 00:37:21}   & 313\quad 00:00:01     \\
 hwb4-11-21           &         79 & \textBF{71\quad 00:00:00}    & \textBF{71\quad 00:00:04}    & 93\quad 00:00:03      \\
 hwb4-11-23           &         81 & \textBF{75\quad 00:00:00}    & \textBF{75\quad 00:00:03}    & 94\quad 00:00:04      \\
 hwb4tc               &        262 & 201\quad 00:05:56       & \textBF{189\quad 00:56:30}   & 267\quad 00:00:05     \\
 hwb5-24-102          &        422 & 348\quad 00:00:14       & \textBF{299\quad 00:56:48}   & 403\quad 00:00:01     \\
 hwb5-24-114          &        421 & 317\quad 00:08:40       & \textBF{312\quad 00:37:52}   & 370\quad 00:00:05     \\
 hwb5-31-91           &        342 & 298\quad 00:00:00       & \textBF{273\quad 00:25:51}   & 355\quad 00:00:08     \\
 hwb5tc               &       1231 & 1127\quad 01:14:11      & \textBF{1063\quad 00:26:30}  & 1183\quad 00:00:15    \\
 hwb6-42-150          &        549 & 511\quad 00:00:01       & \textBF{494\quad 00:47:41}   & 607\quad 00:00:02     \\
 hwb6-47-107          &        364 & 332\quad 00:00:00       & \textBF{319\quad 00:39:33}   & 470\quad 00:00:06     \\
 hwb6tc               &       5471 & 5109\quad 00:00:14      & \textBF{4654\quad 01:26:29}  & 6837\quad 00:00:53    \\
 hwb7-236             &      13574 & \textBF{12660\quad 00:01:02} & \textBF{12660\quad 00:07:15} & 18828\quad 00:05:39   \\
 hwb7-331-2609a       &       8301 & \textBF{7709\quad 00:00:43}  & \textBF{7709\quad 00:03:38}  & 17374\quad 00:00:29   \\
 hwb7tc               &      17242 & \textBF{16051\quad 00:01:20} & \textBF{16051\quad 00:02:06} & 26157\quad 00:02:43   \\
 hwb8-2710-6940       &      20742 & \textBF{19402\quad 00:01:39} & \textBF{19402\quad 00:02:40} & 45855\quad 00:06:31   \\
 mod5-adder-15         &        347 & 323\quad 00:00:00       & \textBF{270\quad 00:20:26}   & 310\quad 00:09:36     \\
 mod5-adder-17-81      &        306 & 286\quad 00:00:00       & \textBF{243\quad 00:31:32}   & 288\quad 00:00:00     \\
 mod5-adders           &        447 & 420\quad 00:00:00       & \textBF{332\quad 00:44:35}   & 360\quad 00:00:01     \\
 mod5d1               &        100 & 91\quad 00:00:00        & 52\quad 00:17:46        & \textBF{39\quad 00:00:01}  \\
 mod5d2               &         89 & 78\quad 00:00:00        & \textBF{45\quad 00:17:04}    & \textBF{45\quad 00:00:02}  \\
 mod5d4               &         42 & 37\quad 00:00:00        & 37\quad 00:26:19        & \textBF{30\quad 00:00:00}  \\
 mod5mils             &         56 & 53\quad 00:00:00        & 44\quad 00:24:33        & \textBF{34\quad 00:00:00}  \\
 mspk-4-49-12         &        110 & \textBF{104\quad 00:00:00}   & \textBF{104\quad 00:45:52}   & 118\quad 00:00:00     \\
 mspk-4-49-13         &        116 & \textBF{104\quad 00:04:20}   & \textBF{104\quad 00:54:47}   & 112\quad 00:00:00     \\
 mspk-4-49-14         &        118 & \textBF{101\quad 00:13:12}   & \textBF{101\quad 00:34:30}   & 116\quad 00:00:00     \\
 mspk-4b15g-1         &        157 & \textBF{143\quad 00:00:00}   & \textBF{143\quad 01:07:05}   & 166\quad 00:00:00     \\
 mspk-4b15g-2         &        119 & \textBF{105\quad 00:00:00}   & \textBF{105\quad 00:10:48}   & 138\quad 00:00:00     \\
 mspk-4b15g-3         &        126 & \textBF{107\quad 00:19:49}   & \textBF{107\quad 00:43:48}   & 130\quad 00:00:00     \\
 mspk-4b15g-4         &        142 & 117\quad 00:05:27       & 117\quad 00:53:16       & \textBF{113\quad 00:00:00} \\
 mspk-4b15g-5         &        135 & \textBF{112\quad 00:35:29}   & \textBF{112\quad 00:37:09}   & 138\quad 00:00:01     \\
 mspk-hwb4-12         &         83 & \textBF{71\quad 00:00:19}    & \textBF{71\quad 00:01:50}    & 90\quad 00:00:02      \\
 mspk-hwb4-13         &         87 & \textBF{77\quad 00:00:00}    & \textBF{77\quad 00:00:26}    & 85\quad 00:00:03      \\
 mspk-nth-primes4-11  &        227 & 204\quad 00:51:47       & \textBF{171\quad 00:58:29}   & 197\quad 00:00:00     \\
 mspk-nth-primes4-12  &        135 & \textBF{124\quad 00:00:00}   & \textBF{124\quad 00:09:01}   & 142\quad 00:00:00     \\
 mspk-nth-primes4-13  &        133 & \textBF{110\quad 00:23:42}   & \textBF{110\quad 00:29:58}   & 129\quad 00:00:00     \\
 mspk-nth-primes4-14  &        122 & \textBF{106\quad 00:00:05}   & \textBF{106\quad 00:06:59}   & 114\quad 00:00:01     \\
 or5d1                &        157 & 144\quad 00:00:00       & \textBF{123\quad 00:12:36}   & 125\quad 00:00:00     \\
 or5d2                &        209 & \textBF{195\quad 00:00:00}   & \textBF{195\quad 00:08:44}   & 271\quad 00:00:02     \\
 rd53-16-67           &        253 & 233\quad 00:00:00       & \textBF{182\quad 00:31:43}   & 200\quad 00:00:04     \\
 rd53d15              &        416 & 379\quad 00:00:00       & \textBF{271\quad 00:46:59}   & 278\quad 00:00:07     \\
 rd53d15s             &        406 & 373\quad 00:00:00       & \textBF{260\quad 00:37:14}   & 337\quad 00:00:01     \\
 rd53d1               &        493 & 443\quad 00:00:00       & 274\quad 00:08:48       & \textBF{246\quad 00:00:00} \\
 rd53d1mils           &        306 & 278\quad 00:00:00       & \textBF{218\quad 00:10:10}   & 260\quad 00:00:02     \\
 rd53d2               &        187 & 171\quad 00:00:00       & \textBF{164\quad 00:32:13}   & 179\quad 00:00:03     \\
 rd53rcmg             &        869 & 808\quad 00:00:02       & \textBF{590\quad 00:14:22}   & 735\quad 00:00:01     \\
 rd73d2               &        301 & 277\quad 00:00:00       & \textBF{266\quad 01:03:13}   & 370\quad 00:00:03     \\
 rd84d1               &        441 & \textBF{410\quad 00:00:01}   & \textBF{410\quad 00:09:09}   & 644\quad 00:00:04     \\
 t6 1 52              &        253 & 238\quad 00:00:00       & \textBF{199\quad 00:41:57}   & 204\quad 00:00:06     \\
 t6 3 48              &        264 & 231\quad 00:00:00       & 175\quad 00:15:10       & \textBF{168\quad 00:00:00} \\
 t7 1 84              &        335 & 312\quad 00:00:00       & 230\quad 00:16:03       & \textBF{225\quad 00:00:01} \\
 t7 4 64              &        349 & 304\quad 00:00:00       & 228\quad 00:27:00       & \textBF{214\quad 00:00:02} \\
 t8 1 116             &        416 & 387\quad 00:00:00       & \textBF{315\quad 00:21:58}   & 319\quad 00:00:03     \\
 t8 5 80              &        434 & 377\quad 00:00:01       & 281\quad 00:21:51       & \textBF{276\quad 00:00:00} \\
\hline
\end{longtable}

\subsection*{Two-Qubit Gate Count}

\begin{figure}[h!]
  \begin{center}
    \includegraphics{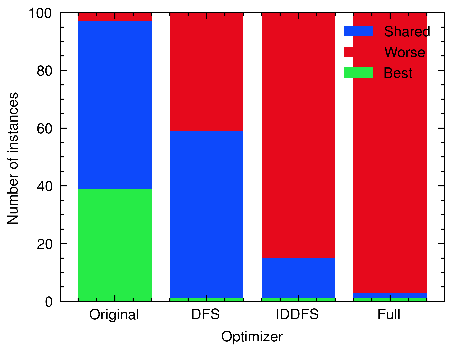}
    \caption{Performance comparison of the best solution found.}
  \end{center}
\end{figure}
\begin{longtable}{lrrrrr}
\hline
 Circuit              & Original   & DFS                 & IDDFS               & Full reduce       \\
\hline
 2of5d1               & \textBF{162}    & \textBF{162\quad 00:00:01}   & 194\quad 00:11:14       & 188\quad 00:00:00     \\
 2of5d1s              & \textBF{126}    & 134\quad 00:00:00       & 177\quad 00:11:25       & 164\quad 00:00:02     \\
 2of5d2               & \textBF{39}     & \textBF{39\quad 00:00:00}    & \textBF{39\quad 00:26:09}    & 86\quad 00:00:03      \\
 2of5d3               & \textBF{99}     & \textBF{99\quad 00:00:00}    & 117\quad 00:12:37       & 127\quad 00:00:03     \\
 3-17tc               & \textBF{15}     & 26\quad 00:00:29        & 26\quad 00:00:46        & 20\quad 00:00:00      \\
 4-49-12-32           & \textBF{35}     & 38\quad 00:00:00        & 38\quad 00:39:35        & 55\quad 00:00:00      \\
 4-49tc1              & \textBF{74}     & 95\quad 01:12:23        & 105\quad 01:17:53       & 97\quad 00:00:00      \\
 4b15g-1              & \textBF{54}     & 73\quad 00:56:23        & 71\quad 00:58:01        & 76\quad 00:00:01      \\
 4b15g-2              & \textBF{71}     & 74\quad 00:00:00        & 100\quad 00:40:59       & 93\quad 00:00:01      \\
 4b15g-3              & \textBF{66}     & 110\quad 00:04:10       & 87\quad 00:57:07        & 94\quad 00:00:02      \\
 4b15g-4              & \textBF{54}     & 73\quad 00:46:26        & 76\quad 00:48:00        & 77\quad 00:00:04      \\
 4b15g-5              & \textBF{47}     & 69\quad 00:18:48        & 69\quad 00:22:16        & 61\quad 00:00:05      \\
 5bit-adder            & \textBF{74}     & 95\quad 00:00:00        & 87\quad 00:34:18        & 101\quad 00:00:00     \\
 5mod5-10-71a         & \textBF{57}     & \textBF{57\quad 00:00:00}    & 108\quad 00:42:36       & 103\quad 00:00:00     \\
 5mod5-8              & \textBF{67}     & \textBF{67\quad 00:00:00}    & 112\quad 00:28:34       & 98\quad 00:00:01      \\
 5mod5-fc             & \textBF{78}     & \textBF{78\quad 00:00:00}    & 121\quad 00:27:40       & 123\quad 00:00:02     \\
 5mod5tc              & \textBF{185}    & \textBF{185\quad 00:00:00}   & 192\quad 00:25:56       & \textBF{185\quad 00:00:00} \\
 6symd2               & \textBF{75}     & \textBF{75\quad 00:00:00}    & \textBF{75\quad 00:15:21}    & 157\quad 00:00:01     \\
 9symd2               & \textBF{114}    & 120\quad 00:00:00       & 120\quad 00:25:54       & 248\quad 00:00:02     \\
 gf$2^{10}$-mult-109-509   & \textBF{609}    & \textBF{609\quad 00:00:02}   & 1751\quad 01:12:09      & 2420\quad 00:00:08    \\
 gf$2^{11}$-mult-131-615   & \textBF{736}    & \textBF{736\quad 00:00:04}   & 2167\quad 00:55:05      & 2433\quad 00:00:08    \\
 gf$2^{12}$-mult-177-753   & \textBF{897}    & \textBF{897\quad 00:00:21}   & 3056\quad 00:38:58      & 3932\quad 00:00:39    \\
 gf$2^{13}$-mult-205-881   & \textBF{1050}   & \textBF{1050\quad 00:00:08}  & 3578\quad 00:51:09      & 4622\quad 00:01:31    \\
 gf$2^{14}$-mult-235-1019  & \textBF{1215}   & \textBF{1215\quad 00:00:11}  & 4398\quad 00:38:27      & 5399\quad 00:02:27    \\
 gf$2^{15}$-mult-239-1139  & \textBF{1364}   & \textBF{1364\quad 00:00:48}  & 4534\quad 00:45:25      & 5856\quad 00:00:21    \\
 gf$2^{16}$-mult-301-1325  & \textBF{1581}   & \textBF{1581\quad 00:03:52}  & 5791\quad 01:07:04      & 6623\quad 00:01:38    \\
 gf$2^{17}$-mult-305-1461  & \textBF{1750}   & \textBF{1750\quad 00:01:07}  & 6239\quad 00:58:45      & 7717\quad 00:03:24    \\
 gf$2^{18}$-mult-375-1671  & \textBF{1995}   & \textBF{1995\quad 00:01:37}  & 7533\quad 00:44:43      & 9110\quad 00:05:23    \\
 gf$2^{19}$-mult-415-1859  & \textBF{2220}   & \textBF{2220\quad 00:01:58}  & 8386\quad 00:50:54      & 8394\quad 00:00:55    \\
 gf$2^{20}$-mult-419-2019  & \textBF{2419}   & \textBF{2419\quad 00:09:43}  & 8585\quad 01:13:34      & 11902\quad 00:04:11   \\
 gf$2^{32}$-mult-1117-5213 & \textBF{6237}   & \textBF{6237\quad 01:24:51}  & \textBF{6237\quad 01:21:55}  & 35491\quad 00:14:33   \\
 gf$2^{32}$-mult-1148-5244 & \textBF{6268}   & \textBF{6268\quad 01:20:04}  & \textBF{6268\quad 01:18:24}  & 27651\quad 00:41:38   \\
 gf$2^{32}$-mult-1179-5275 & \textBF{6299}   & \textBF{6299\quad 00:26:01}  & \textBF{6299\quad 00:28:02}  & 28205\quad 00:08:39   \\
 gf$2^{3}$-mult-11-47      & \textBF{56}     & \textBF{56\quad 00:00:00}    & 117\quad 00:43:42       & 147\quad 00:31:21     \\
 gf$2^{4}$-mult-19-83      & \textBF{99}     & \textBF{99\quad 00:00:00}    & 255\quad 00:33:04       & 310\quad 00:31:22     \\
 gf$2^{50}$-2647-12647    & \textBF{15147}  & \textBF{15147\quad 00:27:15} & \textBF{15147\quad 00:22:51} & 100354\quad 01:29:21  \\
 gf$2^{5}$-mult-29-129     & \textBF{154}    & \textBF{154\quad 00:00:01}   & 390\quad 00:49:11       & 410\quad 00:00:00     \\
 gf$2^{6}$-mult-41-185     & \textBF{221}    & \textBF{221\quad 00:00:02}   & 565\quad 00:39:17       & 820\quad 08:36:33     \\
 gf$2^{7}$-mult-55-251     & \textBF{300}    & \textBF{300\quad 00:00:03}   & 814\quad 00:35:02       & 1074\quad 08:36:36    \\
 gf$2^{8}$-mult-85-341     & \textBF{405}    & \textBF{405\quad 00:00:14}   & 1150\quad 00:18:54      & 1530\quad 00:00:02    \\
 gf$2^{9}$-mult-89-413     & \textBF{494}    & \textBF{494\quad 00:00:07}   & 1353\quad 00:24:47      & 1852\quad 00:00:11    \\
 graycode6            & \textBF{5}      & \textBF{5\quad 00:00:00}     & \textBF{5\quad 00:19:40}     & \textBF{5\quad 00:00:18}   \\
 ham15-109-214        & \textBF{236}    & \textBF{236\quad 00:00:02}   & 308\quad 00:41:51       & 369\quad 00:00:19     \\
 ham15-70             & 534        & \textBF{498\quad 00:30:22}   & 509\quad 00:31:46       & 598\quad 00:00:03     \\
 ham3tc               & \textBF{10}     & 15\quad 00:00:00        & 15\quad 00:00:00        & 15\quad 00:00:10      \\
 ham7-21-69           & \textBF{80}     & 127\quad 01:26:26       & 112\quad 00:29:02       & 135\quad 00:00:10     \\
 ham7-25-49           & \textBF{55}     & 91\quad 00:03:11        & 91\quad 01:13:55        & 93\quad 00:00:11      \\
 ham7tc               & \textBF{106}    & 127\quad 01:00:00       & 137\quad 00:37:21       & 132\quad 00:00:01     \\
 hwb4-11-21           & \textBF{24}     & 29\quad 00:00:00        & 29\quad 00:00:04        & 38\quad 00:00:03      \\
 hwb4-11-23           & \textBF{26}     & 35\quad 00:00:00        & 35\quad 00:00:03        & 31\quad 00:00:04      \\
 hwb4tc               & \textBF{81}     & 88\quad 00:05:56        & 91\quad 00:56:30        & 109\quad 00:00:05     \\
 hwb5-24-102          & \textBF{132}    & 147\quad 00:00:14       & 155\quad 00:56:48       & 162\quad 00:00:01     \\
 hwb5-24-114          & \textBF{131}    & 153\quad 00:08:40       & 153\quad 00:37:52       & 161\quad 00:00:05     \\
 hwb5-31-91           & \textBF{108}    & 119\quad 00:00:00       & 133\quad 00:25:51       & 159\quad 00:00:08     \\
 hwb5tc               & \textBF{386}    & 493\quad 01:14:11       & 466\quad 00:26:30       & 464\quad 00:00:15     \\
 hwb6-42-150          & \textBF{175}    & \textBF{175\quad 00:00:01}   & 256\quad 00:47:41       & 242\quad 00:00:02     \\
 hwb6-47-107          & \textBF{116}    & 137\quad 00:00:00       & 179\quad 00:39:33       & 177\quad 00:00:06     \\
 hwb6tc               & \textBF{1705}   & \textBF{1705\quad 00:00:14}  & 1988\quad 01:26:29      & 1863\quad 00:00:53    \\
 hwb7-236             & \textBF{4222}   & \textBF{4222\quad 00:01:02}  & \textBF{4222\quad 00:07:15}  & 4685\quad 00:05:39    \\
 hwb7-331-2609a       & \textBF{2638}   & \textBF{2638\quad 00:00:43}  & \textBF{2638\quad 00:03:38}  & 3543\quad 00:00:29    \\
 hwb7tc               & \textBF{5355}   & \textBF{5355\quad 00:01:20}  & \textBF{5355\quad 00:02:06}  & 5753\quad 00:02:43    \\
 hwb8-2710-6940       & \textBF{7073}   & \textBF{7073\quad 00:01:39}  & \textBF{7073\quad 00:02:40}  & 9578\quad 00:06:31    \\
 mod5-adder-15         & \textBF{102}    & \textBF{102\quad 00:00:00}   & 119\quad 00:20:26       & 142\quad 00:09:36     \\
 mod5-adder-17-81      & \textBF{90}     & \textBF{90\quad 00:00:00}    & 114\quad 00:31:32       & 107\quad 00:00:00     \\
 mod5-adders           & \textBF{133}    & \textBF{133\quad 00:00:00}   & 183\quad 00:44:35       & 195\quad 00:00:01     \\
 mod5d1               & 28         & 28\quad 00:00:00        & 27\quad 00:17:46        & \textBF{22\quad 00:00:01}  \\
 mod5d2               & \textBF{21}     & \textBF{21\quad 00:00:00}    & 22\quad 00:17:04        & 28\quad 00:00:02      \\
 mod5d4               & \textBF{8}      & \textBF{8\quad 00:00:00}     & \textBF{8\quad 00:26:19}     & 16\quad 00:00:00      \\
 mod5mils             & \textBF{14}     & \textBF{14\quad 00:00:00}    & 26\quad 00:24:33        & 16\quad 00:00:00      \\
 mspk-4-49-12         & \textBF{35}     & 38\quad 00:00:00        & 38\quad 00:45:52        & 59\quad 00:00:00      \\
 mspk-4-49-13         & \textBF{37}     & 52\quad 00:04:20        & 52\quad 00:54:47        & 55\quad 00:00:00      \\
 mspk-4-49-14         & \textBF{38}     & 55\quad 00:13:12        & 55\quad 00:34:30        & 59\quad 00:00:00      \\
 mspk-4b15g-1         & \textBF{47}     & 61\quad 00:00:00        & 61\quad 01:07:05        & 63\quad 00:00:00      \\
 mspk-4b15g-2         & \textBF{37}     & 42\quad 00:00:00        & 42\quad 00:10:48        & 63\quad 00:00:00      \\
 mspk-4b15g-3         & \textBF{38}     & 49\quad 00:19:49        & 49\quad 00:43:48        & 58\quad 00:00:00      \\
 mspk-4b15g-4         & \textBF{44}     & 70\quad 00:05:27        & 70\quad 00:53:16        & 67\quad 00:00:00      \\
 mspk-4b15g-5         & \textBF{40}     & 46\quad 00:35:29        & 46\quad 00:37:09        & 60\quad 00:00:01      \\
 mspk-hwb4-12         & \textBF{25}     & 27\quad 00:00:19        & 27\quad 00:01:50        & 46\quad 00:00:02      \\
 mspk-hwb4-13         & \textBF{28}     & 33\quad 00:00:00        & 33\quad 00:00:26        & 46\quad 00:00:03      \\
 mspk-nth-primes4-11  & \textBF{70}     & 100\quad 00:51:47       & 97\quad 00:58:29        & 88\quad 00:00:00      \\
 mspk-nth-primes4-12  & \textBF{42}     & 49\quad 00:00:00        & 49\quad 00:09:01        & 58\quad 00:00:00      \\
 mspk-nth-primes4-13  & \textBF{43}     & 58\quad 00:23:42        & 58\quad 00:29:58        & 56\quad 00:00:00      \\
 mspk-nth-primes4-14  & \textBF{39}     & 47\quad 00:00:05        & 47\quad 00:06:59        & 52\quad 00:00:01      \\
 or5d1                & \textBF{42}     & 44\quad 00:00:00        & 76\quad 00:12:36        & 70\quad 00:00:00      \\
 or5d2                & \textBF{60}     & \textBF{60\quad 00:00:00}    & \textBF{60\quad 00:08:44}    & 121\quad 00:00:02     \\
 rd53-16-67           & \textBF{75}     & \textBF{75\quad 00:00:00}    & 99\quad 00:31:43        & 97\quad 00:00:04      \\
 rd53d15              & \textBF{118}    & \textBF{118\quad 00:00:00}   & 170\quad 00:46:59       & 163\quad 00:00:07     \\
 rd53d15s             & \textBF{120}    & \textBF{120\quad 00:00:00}   & 181\quad 00:37:14       & 155\quad 00:00:01     \\
 rd53d1               & 144        & 156\quad 00:00:00       & \textBF{138\quad 00:08:48}   & 157\quad 00:00:00     \\
 rd53d1mils           & \textBF{92}     & 96\quad 00:00:00        & 142\quad 00:10:10       & 133\quad 00:00:02     \\
 rd53d2               & \textBF{44}     & \textBF{44\quad 00:00:00}    & 78\quad 00:32:13        & 85\quad 00:00:03      \\
 rd53rcmg             & \textBF{258}    & \textBF{258\quad 00:00:02}   & 395\quad 00:14:22       & 344\quad 00:00:01     \\
 rd73d2               & \textBF{78}     & \textBF{78\quad 00:00:00}    & 81\quad 01:03:13        & 201\quad 00:00:03     \\
 rd84d1               & \textBF{119}    & \textBF{119\quad 00:00:01}   & \textBF{119\quad 00:09:09}   & 275\quad 00:00:04     \\
 t6 1 52              & \textBF{72}     & \textBF{72\quad 00:00:00}    & 114\quad 00:41:57       & 124\quad 00:00:06     \\
 t6 3 48              & \textBF{72}     & \textBF{72\quad 00:00:00}    & 93\quad 00:15:10        & 91\quad 00:00:00      \\
 t7 1 84              & \textBF{96}     & \textBF{96\quad 00:00:00}    & 133\quad 00:16:03       & 133\quad 00:00:01     \\
 t7 4 64              & \textBF{96}     & \textBF{96\quad 00:00:00}    & 131\quad 00:27:00       & 112\quad 00:00:02     \\
 t8 1 116             & \textBF{120}    & \textBF{120\quad 00:00:00}   & 180\quad 00:21:58       & 186\quad 00:00:03     \\
 t8 5 80              & \textBF{120}    & \textBF{120\quad 00:00:01}   & 159\quad 00:21:51       & 160\quad 00:00:00     \\
\hline
\end{longtable}

\end{document}